\documentclass{aa}
\usepackage{graphics}
\usepackage{times}
\begin{document}
\thesaurus{12(12.03.4; 12.04.1; 12.04.3; 11.04.1;
11.07.1; 11.11.1)}
\title{
Investigations of the Local Supercluster Velocity Field
}
\subtitle{
III. Tracing the backside infall with distance moduli from the
direct Tully-Fisher relation 
}
\author{
T.~Ekholm\inst{1,2}
\and
P.~Lanoix\inst{1}
\and
P.~Teerikorpi\inst{2}
\and
P.~Fouqu\'e\inst{3}
\and
G.~Paturel\inst{1}
}
\offprints{
T.~Ekholm
}
\institute{
CRAL - Observatoire de Lyon,
F69561 Saint Genis Laval CEDEX, France
\and
Tuorla Observatory,
FIN-21500 Piikki\"o,
Finland
\and
ESO, Santiago, Chile
}
\date{received, accepted}
\maketitle
%
%
%
%
%
\begin{abstract}
We have extended the discussion of Paper II (Ekholm et al.
\cite{Ekholm99a}) to cover also 
the backside of the Local Supercluster (LSC) by using 
96 galaxies within $\Theta<30\degr$ from the
adopted centre of LSC and with distance moduli from the
direct B-band Tully-Fisher relation. In order to
minimize the influence of the Malmquist bias we required
$\log V_\mathrm{max}>2.1$ and $\sigma_{B_T}<0.2^\mathrm{mag}$.

We found out that if 
$R_\mathrm{Virgo}<20\mathrm{\ Mpc}$ this sample
fails to follow the expected dynamical pattern
from the Tolman-Bondi (TB) model. When
we compared our results with the Virgo core galaxies
given by Federspiel et al. (\cite{Federspiel98}) we
were able to constrain the distance to Virgo:
$R_\mathrm{Virgo}=20-24\mathrm{\ Mpc}$. 

When analyzing the TB-behaviour of the sample
as seen from the origin of the metric as well as
that with distances from the extragalactic Cepheid 
$PL$-relation we found additional support to the estimate
$R_\mathrm{Virgo}=21\mathrm{\ Mpc}$ given in
Paper II. Using a
two-component mass-model we found a Virgo mass
estimate
$M_\mathrm{Virgo}=(1.5$ -- $2)\times M_\mathrm{virial}$, where
$M_\mathrm{virial}=9.375\times10^{14}M_{\sun}$ for
$R_\mathrm{Virgo}=21\mathrm{\ Mpc}$.
This estimate agrees with the conclusion in
Paper I (Teerikorpi et al. \cite{Teerikorpi92}).

Our results indicate that the density distribution of
luminous matter is shallower than that of the total gravitating
matter when $q_0\leq 0.5$. 
The preferred exponent in the density power law,
$\alpha\approx2.5$, agrees with recent theoretical work on the
universal density profile of dark matter clustering
in an Einstein-deSitter universe
(Tittley \& Couchman \cite{Tittley99}).

\end{abstract}
%
%
%
%
%
\section{Introduction}
Study of the local extragalactic velocity field has a
considerable history.  Rubin (\cite{Rubin88}) pinpoints the beginning
of the studies concerning deviations from the Hubble law to
a paper of Gamow (\cite{Gamow46}) where Gamow asked if galaxies partake of
a large-scale systematic rotation in addition to the Hubble
expansion.  The pioneer works by Rubin (\cite{Rubin51}) and Ogorodnikov
(\cite{Ogorodnikov52}) 
gave evidence that the local extragalactic velocity field
is neither linear nor isotropic.
De Vaucouleurs (\cite{deVaucouleurs53}) 
then interpreted the distribution of bright
galaxies and proposed rotation in terms of a flattened local
supergalaxy. 
This short but remarkable paper did not yet refer
to differential expansion, introduced by 
de Vaucouleurs (\cite{deVaucouleurs58})
as an explanation of the ``north-south anisotropy" which he
stated was first pointed out by Sandage (Humason et al. \cite{Humason56}).
Differential expansion was a milder form of Hubble's ``the law of
redshifts does not operate within the Local Group" and de
Vaucouleurs pondered that ``in condensed regions of space, such
as groups or clusters, the expansion rate is greatly
reduced...".
Though there was a period of debate on the importance of the
kinematic effects claimed by de Vaucouleurs and even on the
reality of the local supergalaxy (presently termed as the Local
Supercluster, LSC), already for two decades the reality of the
differential peculiar velocity field around the Virgo cluster
has been generally accepted.
However, its amplitude and such details as the deviation from
spherical symmetry and possible rotational component, are still
under discussion.

A theoretical line of research related to de Vaucouleurs'
differential expansion, has been motivated by the work on
density perturbations in Friedmann cosmological models,
resulting in infall models of matter (Silk \cite{Silk74}) 
which predict a
connection between the infall peculiar velocity at the position
of the Local Group towards the Virgo cluster and the density
parameter of the Friedmann universe. Later on, Olson \& Silk
(\cite{Olson79}) further developed the formalism in a way
which was found useful in Teerikorpi et al.
(\cite{Teerikorpi92}; hereafter Paper I).
The linearized approximation
of Peebles (\cite{Peebles76}) has been often used for describing the
velocity field and for making routine corrections for systemic
velocities.

Using Tolman-Bondi model (Tolman \cite{Tolman34}, Bondi \cite{Bondi47})
Hoffman et al. (\cite{Hoffman80}) calculated the
expected velocity dispersions along line-of-sight as a function
of angular distance from a supercluster and applied the results
to Virgo.  They derived a gravitating mass of about 
$4\times10^{14}M_{\sun}\times100/h_0$
inside the cone of $6\degr$. The Tolman-Bondi (TB) model is
the simplest inhomogeneous
solution to the Einstein's field equations. It describes the time
evolution of a spherically symmetric pressure-free dust
universe in terms of comoving coordinates. For details of the
TB-model cf. 
Ekholm et al. (\cite{Ekholm99a}; hereafter Paper II).

Then, following the course of Hoffman et al. (\cite{Hoffman80}), 
Tully \& Shaya (\cite{Tully84})
calculated the expected run of radial velocity vs.
distance at different angular distances from Virgo and for
different (point) mass-age models.  Comparison of such envelope
curves with available galaxy data agreed with the point
mass having roughly the value of Virgo's virial mass 
($7.5\times10^{14}M_{\sun}\times75/h_0$) 
for reasonable Friedmann universe ages.

The Hubble diagram of Tully \& Shaya contained a small number of
galaxies and did not very well show the expected behaviour.
With a larger sample of Tully-Fisher measured galaxies and attempting to
take into account the Malmquist bias, Teerikorpi et al. 
(\cite{Teerikorpi92})
were able to put in evidence the expected features: an initial
steeply rising tight velocity-distance relation, the local
maximum in front of Virgo and the final ascending part of the
relation, expected to approach asymptotically the undisturbed
Hubble law. Looking from the Virgo centre the zero-velocity surface
was clearly seen around $r/R_\mathrm{Virgo}\approx0.5$.
Using either a continuous mass model or a
two-component model, the conclusions of Tully \& Shaya (\cite{Tully84})
were generally confirmed and it was stated that ``Various density
distributions, constrained by the mass inside the Local Group
distance (required to produce $V_\mathrm{Virgo}$), agree with the
observations, but only if the mass within the Virgo $6\degr$ region
is close to or larger than the standard Virgo virial mass
values.  This is so independently of the value of $q_0$, of the
slope of the density distribution outside of Virgo, and of the
values adopted for Virgo distance and velocity".

It is the aim of the present paper to use the available sample of
galaxies with more accurate distances from Cepheids and
Tully-Fisher relation to study the virgocentric velocity field.
In Paper II
galaxies with
Cepheid-distances were used to map the velocity field in front
of Virgo, here we add galaxies with good Tully-Fisher distances in order
to see both the frontside and backside behaviour and investigate
how conclusions of Paper I should be modified in the light of new
data. It should be emphasized that also our Tully-Fisher distances are now
better, after a programme to study the slope and the Hubble type
dependence of the zero-point (see Theureau et al. \cite{Theureau97}).

This paper is structured as follows. In Sect.~2 we shortly review
the basics of the use of the direct Tully-Fisher relation, give
the relation to be used and describe our sample and the restrictions
put upon it. In Sect.~3 we examine our sample in terms of systemic
velocity vs. distance diagrams and see which distance to Virgo will
bring about best agreement between the TB-predictions and the observations.
In Sect.~4 we try to answer the question whether we have actually
found the Virgo cluster at the centre of the TB-metric.
In Sect.~5 we re-examine our sample from a virgocentric viewpoint
and compare our results from the TF-distances with the sample of
galaxies with distances from the extragalactic Cepheid $PL$-relation.
In Sect.~6 we shortly discuss the mass estimate and our density
profile and, finally, in Sect.~7 we summarize our results with some
conclusive remarks. 
%
%
%
%
\section{The sample based on direct B-band Tully-Fisher relation}
The absolute magnitude $M$ and the logarithm of the maximum
rotational velocity $\log V_\mathrm{max}$ of a galaxy 
(for which also a shorthand $p$ is used) are related as:
%
\begin{equation}
\label{E1}
\\ M=a\log V_\mathrm{max}+b.
\end{equation}
The use of this kind of relation as a distance indicator was suggested
by Gouguenheim (\cite{Gouguenheim69}). Eq.~\ref{E1} is known as
Tully-Fisher (TF) relation after Tully \& Fisher (\cite{Tully77}).

It is nowadays widely acknowledged that the distance moduli
inferred using Eq.~\ref{E1} are underestimated because of
selection effects in the sampling. We can see how this 
Malmquist bias affects the distance determination by
considering the {\it observed} average absolute magnitude 
$\langle M\rangle_p$ at each $p$ 
as a function of the true distance $r$. 
The limit in apparent magnitude, $m_\mathrm{lim}$, 
cuts off progressively more and more of the distribution 
function of $M$ for a constant $p$. This means that the
{\it observed} mean absolute magnitude $\langle M\rangle_p$
is overestimated by the expectation value
$E(M\vert\, p)=ap+b$: 
%
\begin{equation}
\label{E2}
\\ \langle M\rangle_p \le E(M\vert\, p),
\end{equation}

This inequality gives a practical measure of the Malmquist
bias depending primarily on $p$, $r$, $\sigma_M$ 
and $m_\mathrm{lim}$. 
The equality holds only when the magnitude limit cuts the
luminosity function $\Phi(M)$ insignificantly. For our
present purposes it is also important to note that for
luminous galaxies, which are also fast rotators (large $p$) 
the effect of the magnitude limit
is felt at much larger distances than for intrinsically 
faint galaxies which rotate slowly. Hence by limiting
$p$ to large values one expects to add to the sample
galaxies which suffer very little from the
Malmquist bias within a restricted distance range.
For this kind of bias the review by Teerikorpi 
(\cite{Teerikorpi97}) suggested the name Malmquist bias
of the $2^\mathrm{nd}$ kind, in order to make a
difference from the classical Malmquist bias
(of the $1^\mathrm{st}$ kind).

Following Paper I we selected galaxies towards Virgo 
by requiring $\log V_\mathrm{max}$ to be larger than $2.1$. At
the time Paper I was written this value was expected to bring about
nearly unbiased TF distance moduli up to twice the Virgo distance. With
the present, much deeper sample the limit chosen is much safer. 
Also, we allow an error in B-magnitude to be at maximum $0.2^\mathrm{mag}$. 
We also require the axis ratio
to be $\log R_{25}>0.07$. 
Because the maximum amplitude of systemic velocities near Virgo
can be quite large, we first restricted the velocities by
$V_\mathrm{obs}<3V_\mathrm{Virgo}^\mathrm{cosm}\cos\Theta$,
where $\Theta$ is the angular distance from the adopted centre
($l=284\degr$, $b=74.5\degr$) and the cosmological velocity of the
centre is following Paper II 
$V_\mathrm{Virgo}^\mathrm{cosm}=1200\mathrm{\ km\,s^{-1}}$.
After this the derived TF-distances were restricted by
$R_\mathrm{TF}<60\mathrm{\ Mpc}$. 

%
\begin{table*}
\caption{Basic information for the 96 galaxies accepted to our
TF-sample within $30\degr$ from the centre. For an explanation
of the entries cf. Sect.~2. 
}
\begin{center}
\begin{tabular}{clccccccccccc}
\hline
\hline
 PGC & Name & $l$ & $b$ & $T$ & $\log R_{25}$ & $B_\mathrm{T}^\mathrm{c}$ &
 $\sigma_B$ & $\log V_\mathrm{max}$ & $\sigma_{\log V_\mathrm{max}}$ &
 $V_\mathrm{obs}$ & $\Theta_\mathrm{gal}$ & $R_\mathrm{gal}$ \\
\hline
(1) & (2) & (3) & (4) & (5) & (6) & (7) & (8) & (9) & (10) & (11) & (12) & (13) \\
\hline
031883 & NGC  3338 & 230.33 & 57.02 & 5.3 &  .21 & 11.02 & .12 & 2.268 & .033 & 1174 & 26.55 & 28.32 \\
032007 & NGC  3351 & 233.95 & 56.37 & 2.5 &  .23 &  9.97 & .13 & 2.166 & .027 &  640 & 26.18 & 11.52 \\
032192 & NGC  3368 & 234.44 & 57.01 & 1.5 &  .17 &  9.65 & .12 & 2.318 & .039 &  761 & 25.49 & 12.56 \\
032306 & NGC  3389 & 233.72 & 57.74 & 6.1 &  .35 & 11.52 & .10 & 2.116 & .029 & 1168 & 25.04 & 24.55 \\
033166 & NGC  3486 & 202.08 & 65.49 & 5.1 &  .14 & 10.55 & .08 & 2.125 & .050 &  632 & 26.83 & 15.54 \\
033234 & NGC  3495 & 249.89 & 54.73 & 5.9 &  .64 & 11.08 & .08 & 2.233 & .006 &  960 & 23.86 & 27.44 \\
034612 & NGC  3623 & 241.33 & 64.22 & 1.6 &  .61 &  9.36 & .16 & 2.395 & .005 &  676 & 17.60 & 13.51 \\
034695 & NGC  3627 & 241.97 & 64.42 & 2.7 &  .33 &  8.92 & .16 & 2.265 & .020 &  596 & 17.28 &  9.26 \\
034697 & NGC  3628 & 240.86 & 64.78 & 3.3 &  .59 &  9.19 & .16 & 2.349 & .009 &  719 & 17.28 & 13.14 \\
034935 & NGC  3655 & 235.59 & 66.97 & 4.9 &  .18 & 11.75 & .08 & 2.260 & .049 & 1364 & 17.02 & 38.80 \\
035043 & NGC  3666 & 246.40 & 64.18 & 5.2 &  .55 & 11.53 & .07 & 2.108 & .009 &  923 & 16.33 & 23.32 \\
035088 & NGC  3672 & 270.42 & 47.55 & 4.7 &  .31 & 11.28 & .16 & 2.328 & .018 & 1635 & 27.58 & 37.50 \\
035224 & NGC  3684 & 235.98 & 68.07 & 4.7 &  .18 & 11.51 & .10 & 2.120 & .032 & 1053 & 16.12 & 23.86 \\
035268 & NGC  3686 & 235.71 & 68.28 & 4.5 &  .10 & 11.51 & .11 & 2.126 & .055 & 1047 & 16.05 & 24.25 \\
035294 & NGC  3689 & 212.72 & 71.32 & 5.8 &  .19 & 12.43 & .09 & 2.209 & .040 & 2674 & 19.89 & 47.91 \\
035405 & NGC  3701 & 217.69 & 71.30 & 3.8 &  .32 & 12.73 & .08 & 2.120 & .021 & 2732 & 18.70 & 41.22 \\
035440 & NGC  3705 & 252.02 & 63.79 & 2.3 &  .38 & 11.10 & .13 & 2.232 & .022 &  871 & 15.28 & 19.44 \\
036243 & NGC  3810 & 252.94 & 67.22 & 5.6 &  .17 & 10.79 & .09 & 2.246 & .029 &  858 & 12.28 & 24.86 \\
036266 & NGC  3813 & 176.19 & 72.42 & 4.4 &  .29 & 11.52 & .07 & 2.202 & .029 & 1459 & 26.63 & 29.42 \\
038031 & NGC  4045 & 275.98 & 62.27 & 1.7 &  .20 & 12.27 & .11 & 2.249 & .051 & 1806 & 12.55 & 34.88 \\
038150 & NGC  4062 & 185.26 & 78.65 & 5.6 &  .35 & 11.04 & .07 & 2.189 & .011 &  743 & 20.48 & 23.94 \\
038693 & NGC  4145 & 154.27 & 74.62 & 6.6 &  .17 & 11.21 & .08 & 2.106 & .034 & 1032 & 27.89 & 15.65 \\
038749 & NGC  4152 & 260.39 & 75.42 & 4.9 &  .10 & 12.24 & .16 & 2.194 & .099 & 2059 &  6.15 & 40.73 \\
038916 & IC    769 & 269.75 & 72.44 & 3.8 &  .17 & 12.82 & .11 & 2.180 & .046 & 2093 &  4.53 & 50.47 \\
038943 & NGC  $4178^1$ & 271.86 & 71.37 & 6.6 &  .46 & 10.86 & .09 & 2.101 & .037 &  245 &  4.73 & 13.14 \\
038964 & NGC  4180 & 276.79 & 67.94 & 3.2 &  .45 & 12.45 & .18 & 2.301 & .018 & 1935 &  6.95 & 51.85 \\
039025 & NGC  $4189^2$ & 268.37 & 73.72 & 6.1 &  .17 & 11.97 & .19 & 2.221 & .086 & 1994 &  4.34 & 40.03 \\
039028 & NGC  $4192^3$ & 265.44 & 74.96 & 2.4 &  .60 &  9.98 & .09 & 2.377 & .005 & -253 &  4.89 & 17.13 \\
039040 & NGC  $4193^4$ & 268.91 & 73.51 & 4.4 &  .32 & 12.43 & .09 & 2.251 & .018 & 2355 &  4.26 & 51.02 \\
039152 & IC   $3061^5$ & 268.20 & 74.39 & 5.0 &  .74 & 12.90 & .08 & 2.136 & .017 & 2201 &  4.23 & 47.25 \\
039224 & NGC  $4212^6$ & 268.89 & 74.36 & 5.6 &  .20 & 11.21 & .15 & 2.175 & .034 & -198 &  4.05 & 24.93 \\
039246 & NGC  $4216^7$ & 270.45 & 73.74 & 2.2 &  .65 &  9.96 & .15 & 2.410 & .005 &   11 &  3.78 & 18.54 \\
039308 & NGC  $4222^8$ & 270.54 & 73.93 & 6.0 &  .79 & 12.30 & .09 & 2.146 & .007 &  111 &  3.70 & 38.11 \\
039389 & NGC  4235 & 279.18 & 68.47 & 1.0 &  .65 & 11.88 & .09 & 2.153 & .009 & 2263 &  6.22 & 22.53 \\
039393 & NGC  $4237^9$ & 267.21 & 75.76 & 4.9 &  .19 & 11.92 & .13 & 2.158 & .034 &  757 &  4.47 & 31.92 \\
039656 & NGC  4260 & 281.56 & 67.63 & 1.0 &  .30 & 12.17 & .06 & 2.388 & .024 & 1695 &  6.91 & 48.36 \\
039724 & NGC  4274 & 191.40 & 82.62 & 1.4 &  .44 & 10.58 & .17 & 2.357 & .015 &  891 & 17.43 & 21.40 \\ 
039738 & NGC  4273 & 282.53 & 66.96 & 5.5 &  .21 & 11.74 & .09 & 2.243 & .031 & 2228 &  7.56 & 38.19 \\
039886 & NGC  4289 & 284.38 & 65.49 & 4.9 & 1.00 & 12.72 & .09 & 2.232 & .010 & 2381 &  9.01 & 56.26 \\
039907 & NGC  4293 & 262.85 & 78.82 & 1.3 &  .28 & 10.70 & .13 & 2.229 & .022 &  839 &  6.45 & 16.04 \\
039974 & NGC  $4302^{10}$ & 272.52 & 75.68 & 5.3 &  .81 & 10.96 & .14 & 2.236 & .009 & 1005 &  3.17 & 25.28 \\
040033 & NGC  $4307^{11}$ & 280.58 & 70.63 & 3.1 &  .66 & 11.65 & .07 & 2.253 & .012 &  956 &  4.00 & 31.54 \\
040119 & NGC  $4316^{12}$ & 280.72 & 70.95 & 4.7 &  .70 & 12.27 & .08 & 2.159 & .012 & 1119 &  3.68 & 37.60 \\
040153 & NGC  $4321^{13}$ & 271.14 & 76.90 & 4.6 &  .09 &  9.65 & .11 & 2.277 & .083 & 1483 &  3.97 & 15.44 \\
040251 & NGC  $4343^{14}$ & 283.56 & 68.77 & 2.4 &  .53 & 12.26 & .18 & 2.221 & .009 &  867 &  5.73 & 32.21 \\
040284 & NGC  4348 & 289.61 & 58.71 & 4.2 &  .66 & 11.94 & .15 & 2.244 & .007 & 1815 & 15.93 & 39.95 \\
040507 & NGC  $4380^{15}$ & 281.94 & 71.82 & 2.6 &  .24 & 12.03 & .14 & 2.175 & .045 &  839 &  2.75 & 30.48 \\
040566 & IC  3322A$^{16}$ & 284.72 & 69.17 & 6.0 &  .89 & 11.87 & .14 & 2.115 & .012 &  852 &  5.33 & 28.77 \\
040581 & NGC  $4388^{17}$ & 279.12 & 74.34 & 2.9 &  .55 & 10.76 & .11 & 2.329 & .006 & 2401 &  1.32 & 25.67 \\
040621 & UGC  7522 & 287.43 & 65.53 & 5.3 &  .95 & 12.75 & .12 & 2.161 & .013 & 1265 &  9.04 & 47.15 \\
040644 & NGC  $4402^{18}$ & 278.79 & 74.78 & 4.4 &  .51 & 11.54 & .19 & 2.152 & .008 &  120 &  1.41 & 25.97 \\
040692 & NGC  4414 & 174.55 & 83.18 & 5.1 &  .23 & 10.28 & .11 & 2.342 & .033 &  694 & 18.87 & 24.56 \\
040914 & NGC  $4438^{19}$ & 280.35 & 74.83 & 2.9 &  .47 & 10.08 & .16 & 2.231 & .018 &  -36 &  1.02 & 14.43 \\
040988 & NGC  4448 & 195.35 & 84.67 & 2.0 &  .43 & 11.22 & .12 & 2.266 & .009 &  618 & 16.25 & 22.51 \\
\hline
\hline
\end{tabular}
\end{center}
\end{table*}
\addtocounter{table}{-1}\begin{table*}
\caption{continued}
\begin{center}
\begin{tabular}{clccccccccccc}
\hline
\hline
 PGC & Name & $l$ & $b$ & $T$ & $\log R_{25}$ & $B_\mathrm{T}^\mathrm{c}$ &
 $\sigma_B$ & $\log V_\mathrm{max}$ & $\sigma_{\log V_\mathrm{max}}$ &
 $V_\mathrm{obs}$ & $\Theta_\mathrm{gal}$ & $R_\mathrm{gal}$ \\
\hline
(1) & (2) & (3) & (4) & (5) & (6) & (7) & (8) & (9) & (10) & (11) & (12) & (13) \\
\hline
041024 & NGC  $4450^{20}$ & 273.91 & 78.64 & 1.8 &  .12 & 10.48 & .07 & 2.413 & .047 & 1862 &  4.74 & 23.75 \\
041317 & NGC  4480 & 289.67 & 66.58 & 5.2 &  .31 & 12.36 & .06 & 2.226 & .017 & 2288 &  8.13 & 46.90 \\
041517 & NGC  $4501^{21}$ & 282.33 & 76.51 & 4.1 &  .27 &  9.56 & .12 & 2.476 & .018 & 2172 &  2.05 & 24.87 \\
041719 & NGC  $4519^{22}$ & 289.17 & 71.05 & 6.3 &  .09 & 11.96 & .06 & 2.146 & .189 & 1087 &  3.77 & 32.59 \\
041789 & NGC  4527 & 292.60 & 65.18 & 3.4 &  .41 & 10.47 & .13 & 2.264 & .014 & 1571 &  9.76 & 18.86 \\
041812 & NGC  $4535^{23}$ & 290.07 & 70.64 & 4.8 &  .12 & 10.19 & .16 & 2.311 & .062 & 1821 &  4.26 & 21.69 \\
041934 & NGC  $4548^{24}$ & 285.70 & 76.83 & 2.8 &  .08 & 10.58 & .11 & 2.291 & .078 &  379 &  2.37 & 21.34 \\
042038 & NGC  4565 & 230.77 & 86.44 & 3.6 &  .86 &  8.90 & .19 & 2.414 & .004 & 1181 & 13.66 & 15.54 \\
042064 & NGC  4567 & 289.78 & 73.75 & 5.2 &  .14 & 11.58 & .09 & 2.243 & .070 & 2145 &  1.75 & 34.28 \\
042069 & NGC  $4568^{25}$ & 289.82 & 73.73 & 5.2 &  .34 & 10.82 & .09 & 2.274 & .010 & 2134 &  1.77 & 26.25 \\
042089 & NGC  $4569^{26}$ & 288.47 & 75.62 & 2.7 &  .35 &  9.51 & .17 & 2.369 & .021 & -355 &  1.61 & 16.07 \\
042168 & NGC  $4579^{27}$ & 290.40 & 74.35 & 2.2 &  .08 & 10.12 & .09 & 2.471 & .047 & 1399 &  1.72 & 23.50 \\
042319 & NGC  4591 & 294.54 & 68.68 & 3.0 &  .30 & 13.26 & .12 & 2.200 & .027 & 2282 &  6.68 & 57.43 \\
042476 & NGC  4602 & 297.89 & 57.63 & 5.1 &  .47 & 10.99 & .17 & 2.309 & .008 & 2351 & 17.68 & 31.18 \\
042741 & NGC  $4639^{28}$ & 294.30 & 75.98 & 3.3 &  .19 & 11.68 & .07 & 2.254 & .052 &  888 &  3.01 & 32.06 \\
042791 & NGC  4642 & 298.57 & 62.16 & 4.6 &  .53 & 12.32 & .11 & 2.132 & .050 & 2476 & 13.37 & 35.79 \\
042816 & NGC  $4647^{29}$ & 295.75 & 74.34 & 5.4 &  .08 & 11.53 & .14 & 2.127 & .048 & 1298 &  3.15 & 24.54 \\
042833 & NGC  $4651^{30}$ & 293.07 & 79.12 & 5.1 &  .18 & 10.81 & .10 & 2.358 & .031 &  711 &  5.05 & 32.73 \\
042857 & NGC  $4654^{31}$ & 295.43 & 75.89 & 5.4 &  .22 & 10.46 & .10 & 2.218 & .037 &  926 &  3.23 & 19.14 \\
043147 & NGC  4682 & 301.23 & 52.79 & 4.4 &  .32 & 12.37 & .10 & 2.191 & .018 & 2126 & 22.81 & 42.25 \\
043186 & NGC  $4689^{32}$ & 299.08 & 76.61 & 5.0 &  .08 & 11.20 & .10 & 2.148 & .073 & 1508 &  4.30 & 22.30 \\
043254 & NGC  4698 & 300.57 & 71.35 & 1.4 &  .16 & 11.00 & .12 & 2.404 & .054 &  872 &  5.77 & 29.45 \\
043331 & NGC  4701 & 301.54 & 66.25 & 4.6 &  .07 & 12.41 & .08 & 2.106 & .127 &  571 & 10.05 & 34.79 \\
043451 & NGC  4725 & 295.09 & 88.36 & 2.1 &  .18 &  9.57 & .15 & 2.380 & .042 & 1160 & 13.89 & 14.29 \\
043601 & NGC  $4746^{33}$ & 303.39 & 74.95 & 3.9 &  .53 & 12.29 & .06 & 2.208 & .020 & 1667 &  5.10 & 42.62 \\
043784 & NGC  4771 & 304.03 & 64.14 & 4.9 &  .65 & 11.59 & .13 & 2.109 & .013 &  969 & 12.41 & 24.04 \\
043798 & NGC  4772 & 304.15 & 65.03 & 1.0 &  .23 & 11.52 & .15 & 2.394 & .041 &  884 & 11.63 & 36.43 \\
043939 & NGC  4793 & 101.55 & 88.05 & 5.4 &  .27 & 11.57 & .12 & 2.248 & .030 & 2466 & 17.45 & 34.58 \\
044191 & NGC  4818 & 305.21 & 54.32 & 2.8 &  .43 & 11.18 & .10 & 2.136 & .017 &  867 & 21.87 & 18.56 \\
044254 & UGC  8067 & 305.92 & 61.13 & 4.4 &  .73 & 12.75 & .10 & 2.154 & .017 & 2668 & 15.51 & 45.58 \\
044392 & NGC  4845 & 306.74 & 64.40 & 2.3 &  .61 & 11.12 & .16 & 2.292 & .007 & 1073 & 12.70 & 23.05 \\
045170 & NGC  4939 & 308.10 & 52.40 & 3.8 &  .36 & 11.07 & .10 & 2.369 & .020 & 2908 & 24.17 & 37.42 \\
045311 & NGC  4961 &  44.51 & 86.76 & 5.8 &  .17 & 13.41 & .15 & 2.118 & .043 & 2513 & 17.37 & 58.94 \\
045643 & NGC  4995 & 310.78 & 54.76 & 3.0 &  .16 & 11.42 & .12 & 2.350 & .041 & 1578 & 22.38 & 36.80 \\
045749 & NGC  5005 & 101.61 & 79.25 & 3.0 &  .30 &  9.92 & .12 & 2.460 & .009 &  967 & 26.24 & 24.77 \\
045948 & NGC  5033 &  98.06 & 79.45 & 5.1 &  .42 &  9.76 & .16 & 2.341 & .010 &  896 & 26.02 & 19.28 \\
046441 & NGC  5073 & 312.94 & 47.48 & 3.7 &  .75 & 12.08 & .18 & 2.281 & .014 & 2533 & 29.74 & 47.06 \\
046671 & NGC  5112 &  96.06 & 76.76 & 5.3 &  .13 & 12.12 & .09 & 2.112 & .067 & 1003 & 28.67 & 30.93 \\
048130 & NGC  5248 & 335.93 & 68.75 & 4.1 &  .11 & 10.44 & .18 & 2.283 & .083 & 1048 & 16.70 & 22.23 \\
049555 & NGC  5364 & 340.71 & 63.03 & 5.4 &  .18 & 10.56 & .19 & 2.239 & .043 & 1128 & 22.28 & 21.20 \\
050782 & NGC  5506 & 339.15 & 53.81 & 3.7 &  .57 & 11.65 & .19 & 2.174 & .015 & 1679 & 29.79 & 28.98 \\
051233 & NGC  5566 & 349.27 & 58.56 & 1.3 &  .46 & 10.66 & .18 & 2.378 & .040 & 1408 & 28.30 & 23.49 \\
\hline
\hline
\end{tabular}
\end{center}
\end{table*}


With these criteria we found 96 galaxies within $\Theta<30\degr$
tabulated in
Table 1, where 
in columns (1) and (2) we give the PGC number and name (the superscript
after some galaxies will be explained in Sect.~4). In
columns (3) and (4) the galactic coordinates 
$l,b$ in degrees are given. 
In column (5)
we give the morphological type code $T$ and
in column (6) we give the logarithm of the
axis ratio at $25^\mathrm{\ mag}/\sq\arcsec$, $\log R_{25}$. 
The total
B-magnitude corrected according to RC3 (de Vaucouleurs et
al. \cite{deVaucouleurs91})\footnote{Except for galactic extinction
which is adopted from RC2 (de Vaucouleurs et al.
\cite{deVaucouleurs76})} 
and the corresponding weighted mean
error are given in columns (7) and (8). In columns (9) and (10) we
give the logarithm of the maximum rotational velocity $\log V_\mathrm{max}$
with the weighted mean error.
In column (11) we give the
observed velocity $V_\mathrm{obs}$ by which -- as in Paper II -- 
we mean the mean observed
heliocentric velocity corrected
to the centroid of the Local Group according to 
Yahil et al. (\cite{Yahil77}). Finally, in columns (12) and (13) we
have the angular distance $\Theta_\mathrm{gal}$ in degrees 
between a galaxy and the centre and
the distance $R_\mathrm{gal}$ in Mpc from us calculated using the
direct TF-relation given below.
The data in columns (1) -- (11) were extracted from 
the Lyon-Meudon extragalactic database LEDA.

Our direct TF-parameters for the B-band magnitudes were
taken from Theureau et al. (\cite{Theureau97}). The slope
for the relation is $a=-5.823$ and the zero-points 
corrected for the type-effect are given in Table~2.
%
\begin{table}
\caption{Type-corrected zero-points.}
\begin{center}
\begin{tabular}{cc}
\hline
\hline
Hubble type code $T$ & zero-point $b$ \\
\hline
1,2 & $-7.347$ \\
3 & $-7.725$ \\ 
4 & $-8.001$ \\ 
5 & $-8.034$ \\ 
6 & $-8.109$ \\ 
7,8 & $-7.499$ \\
\hline
\end{tabular}
\end{center} 
\end{table}
The calibration of the zero-points was based on a sample of
galaxies with Cepheid distances given in Table~1 in
Theureau et al. (\cite{Theureau97}). This calibration
corresponded to a Hubble constant
$H_0\approx55\mathrm{\ km\, s^{-1}\, Mpc^{-1}}$.

Finally we comment on our notation on velocities. We use systemic
velocity in the same sense as in Paper II, i.e. the systemic
velocity is a combination of the cosmological velocity
and the velocity induced by Virgo with the assumption that
the virgocentric motions dominate. When we refer to
observed systemic velocity we call it $V_\mathrm{obs}$ and when
to model prediction, $V_\mathrm{pred}$. If we make no distinction,
we use $V_\mathrm{sys}$. 
%
%
%
%
\section{The $V_\mathrm{sys}$ vs. $R_\mathrm{gal}$ diagram
for the TF-sample}
In Paper II we found a TB-solution using a simple density
law $\rho(R) = \rho_\mathrm{bg}(1+kR^{-\alpha})$,
which fitted data quite well. Here $R$ is the distance from
the origin of the TB-metric, $\alpha$ is the density gradient
and $k$ the density contrast. Because an Einstein-deSitter
universe was assumed, the background density $\rho_\mathrm{bg}$
equals the critical cosmological density, $\rho_\mathrm{c}$.
The relevant quantity, the mass within a radius $d$, the radius $R$
in units of Virgo distance, was expressed as
$M(d) = M(d)_\mathrm{EdS}\times(1+k'd^{-\alpha})$
(cf. Eq.~9 in Paper II). Here $k'$ is the mass excess
within a sphere having a radius of one Virgo distance. 

Unfortunately, the sample of galaxies with distances
from the extragalactic Cepheid $PL$-relation did not reach
well enough behind the LSC. Our present sample is clearly
deep enough to reveal the backside infall signal. In Paper I
it was well seen how in the front the differences between
different TB-models were not large, in contrast to the background,
where the model predictions progressively deviate from each
other.

In the formalism developed by Ekholm (\cite{Ekholm96})
and adopted in Paper II, the quantity given by Eq.~8 in Paper II,
$A(d,q_0)$, which is needed for solving the development angle, is no
longer an explicit function of $H_0$. There are -- however
-- still rather many free parameters, which we shortly discuss
below:
\begin{enumerate}
\item The deceleration parameter $q_0$. In Paper II we
considered $q_0$ given, restricting our analysis to the
Einstein-deSitter universe ($q_0=0.5$). In Paper I it was
concluded that $q_0$ has a minor influence on the 
$V_\mathrm{pred}$ vs. $R_\mathrm{gal}$ curves and on
the Virgo mass (though it has a large effect on total
mass inside the LG sphere). 
\item The density gradient $\alpha$ and the relative mass
excess at $d=1$, $k'$. We remind that $k'$ in our formalism
does not depend on $\alpha$ but only on the amount by which
the LG's expansion velocity with respect to centre of LSC
has slowed down. In our two-component model (Sect.~5) $k'$
will depend also on $\alpha$.
\item The velocities $V_\mathrm{LG}^\mathrm{in}$,
$V_\mathrm{Virgo}^\mathrm{obs}$ and $V_\mathrm{Virgo}^\mathrm{cosm}$.
As in Papers I and II, we presume Virgo to be at rest with
cosmological background:
$V_\mathrm{Virgo}^\mathrm{cosm}=
V_\mathrm{LG}^\mathrm{in}+V_\mathrm{Virgo}^\mathrm{obs}$.
We feel that
our choices for the infall velocity of the Local Group
$V_\mathrm{LG}^\mathrm{in}=220\mathrm{\ km\, s^{-1}}$
and for the observed velocity of Virgo
$V_\mathrm{Virgo}^\mathrm{obs}=980\mathrm{\ km\, s^{-1}}$
are relatively safe.
\end{enumerate}

We would also like to remind that our solutions in Paper II had
an implicit dependence on the Hubble constant $H_0$, because
we fixed our distance to Virgo kinematically from
$R_\mathrm{Virgo}=V_\mathrm{Virgo}^\mathrm{cosm}/H_0$
by adopting $H_0=57\mathrm{\ km\, s^{-1}\, Mpc^{-1}}$.
This {\it global} value was based on SNe Ia
(Lanoix \cite{Lanoix99}) and agrees also with the more
local results of the KLUN ({\sl Kinematics of the Local
Universe}) project (Theureau et al. \cite{Theureau97};
Ekholm et al. \cite{Ekholm99b}) and with the findings
of Federspiel et al. (\cite{Federspiel98}). Here we allow
the distance of Virgo, or equivalently the Hubble constant
$H_0$, vary keeping the cosmological velocity of Virgo
fixed. This choice is justified because even though the
estimates for $H_0$ have converged to 
$\sim60\mathrm{\ km\, s^{-1}\, Mpc^{-1}}$ the reported
$1\sigma$ errors are not small and the different values are 
still scattered ($50$-$70\mathrm{\ km\, s^{-1}\, Mpc^{-1}}$).  
 
In this section we examine how well the present TB-sample
agrees with the Model 1 of Paper II, which
constitutes of a density excess embedded in a FRW universe with
$q_0=0.5$ and $H_0=57\mathrm{\ km\, s^{-1}\, Mpc^{-1}}$.
The model parameters are $k'=0.606$ and $\alpha=2.85$, which
predict for the Virgo cluster ($\Theta<6\degr$) a mass 
$1.62\times M_\mathrm{virial}$, where $M_\mathrm{virial}$
is the virial mass of the Virgo cluster derived by
Tully \& Shaya (\cite{Tully84}) 
$=7.5\times10^{14}M_{\sun}R_\mathrm{Virgo}/16.8\mathrm{\ Mpc}$.
Because of fixed infall velocity of the Local Group (LG) into
the centre of LSC and because $H_0$ was fixed from external
considerations
the distance to centre of LSC became to be 
$R_\mathrm{Virgo}=21\mathrm{\ Mpc}$. 
For further details of the TB-model
adopted cf. Paper II. Additional discussion can be found in Paper I,
Ekholm \& Teerikorpi (\cite{Ekholm94}) 
and Ekholm (\cite{Ekholm96}).

\begin{figure}
\resizebox{\hsize}{!}{\includegraphics{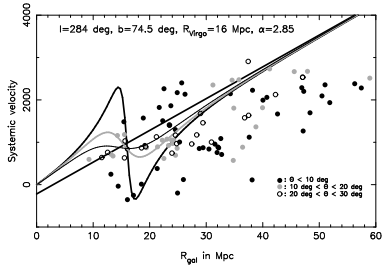}}
\caption{The systemic velocity vs. distance for galaxies listed
in Table 1 for the Model 1 and $R_\mathrm{Virgo}=16\mathrm{\ Mpc}$.
The data points are given with black bullets for
$\Theta<10\degr$, with grey bullets for
$10\degr\le\Theta<20\degr$ and with
circles for $20\degr\le\Theta<30\degr$.
The thick black curve is the theoretical TB-pattern for the
average angular distance $\langle\Theta\rangle=4.5\degr$,
the gray curve is for $\langle\Theta\rangle=15.6\degr$
and the thin black curve for $\langle\Theta\rangle=25.8\degr$.
These values are the mean values of data in each angular
interval.
The straight line is the Hubble law for
$H_0=75\mathrm{\ km\, s^{-1}\, Mpc^{-1}}$ based on the
adopted distance and
$V_\mathrm{Virgo}^\mathrm{cosm}= 1200\mathrm{\ km\, s^{-1}}$.
}
\label{F1}
\end{figure}

\begin{figure}
\resizebox{\hsize}{!}{\includegraphics{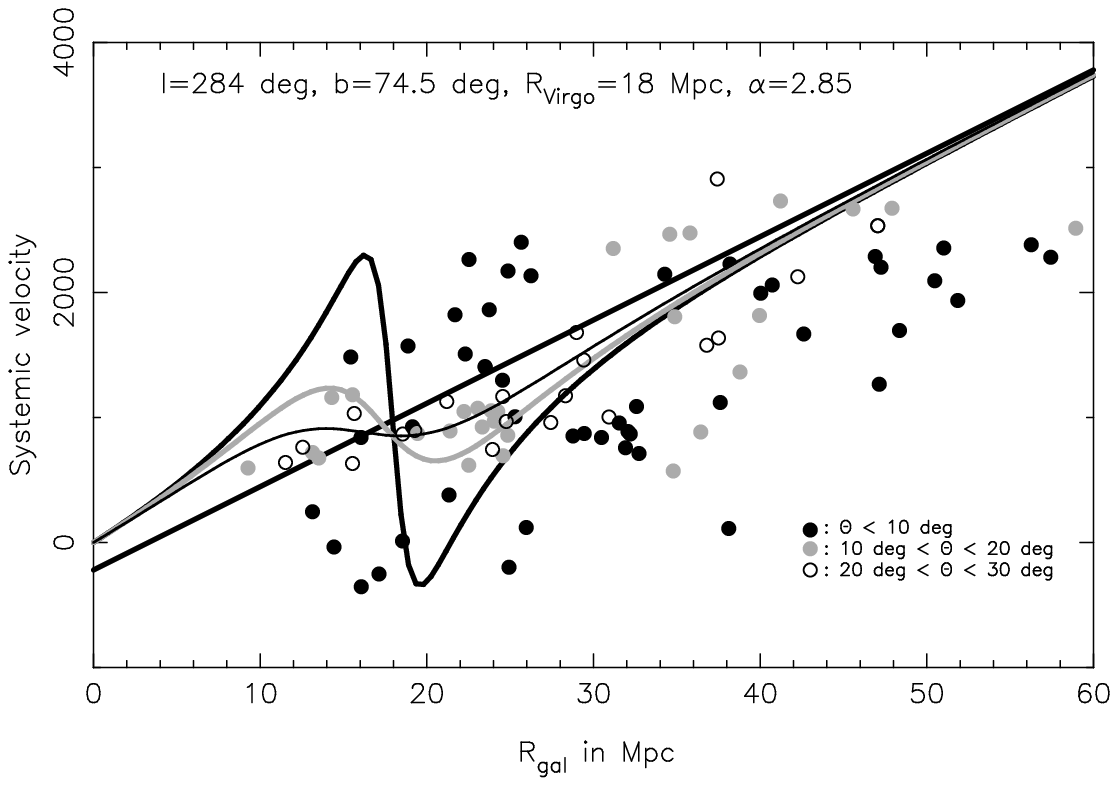}}
\caption{
As Fig.~1, but now the distance to Virgo used for
normalization is $R_\mathrm{Virgo}=18\mathrm{\ Mpc}$,
which corresponds to $H_0=67\mathrm{\ km\, s^{-1}\, Mpc^{-1}}$.
}
\label{F2}
\end{figure}

\begin{figure}
\resizebox{\hsize}{!}{\includegraphics{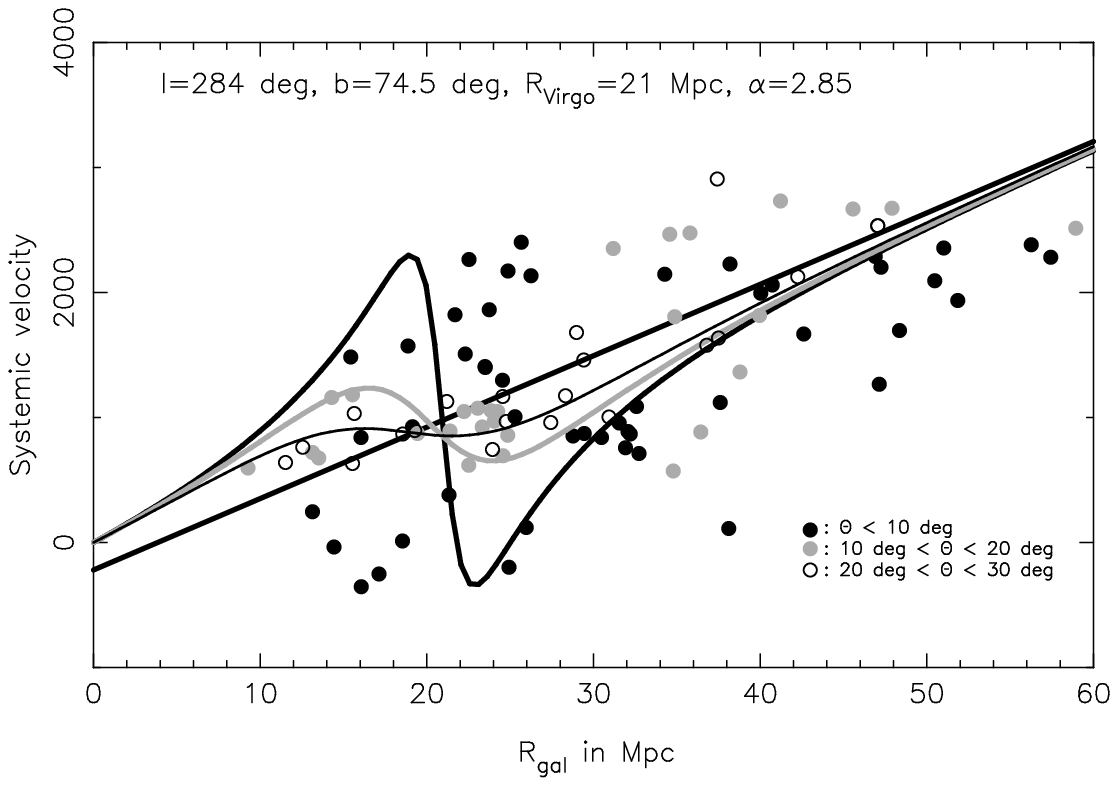}}
\caption{
As Fig.~1, but now the distance to Virgo used for
normalization is $R_\mathrm{Virgo}=21\mathrm{\ Mpc}$,
which corresponds to $H_0=57\mathrm{\ km\, s^{-1}\, Mpc^{-1}}$.
}
\label{F3}
\end{figure}

\begin{figure}
\resizebox{\hsize}{!}{\includegraphics{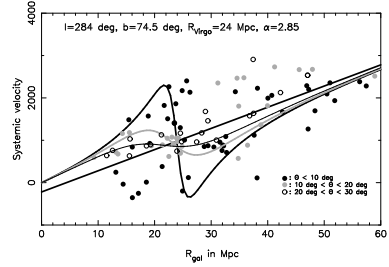}}
\caption{
As Fig.~1, but now the distance to Virgo used for
normalization is $R_\mathrm{Virgo}=24\mathrm{\ Mpc}$,
which corresponds to $H_0=50\mathrm{\ km\, s^{-1}\, Mpc^{-1}}$.
}
\label{F4}
\end{figure}

\begin{figure}
\resizebox{\hsize}{!}{\includegraphics{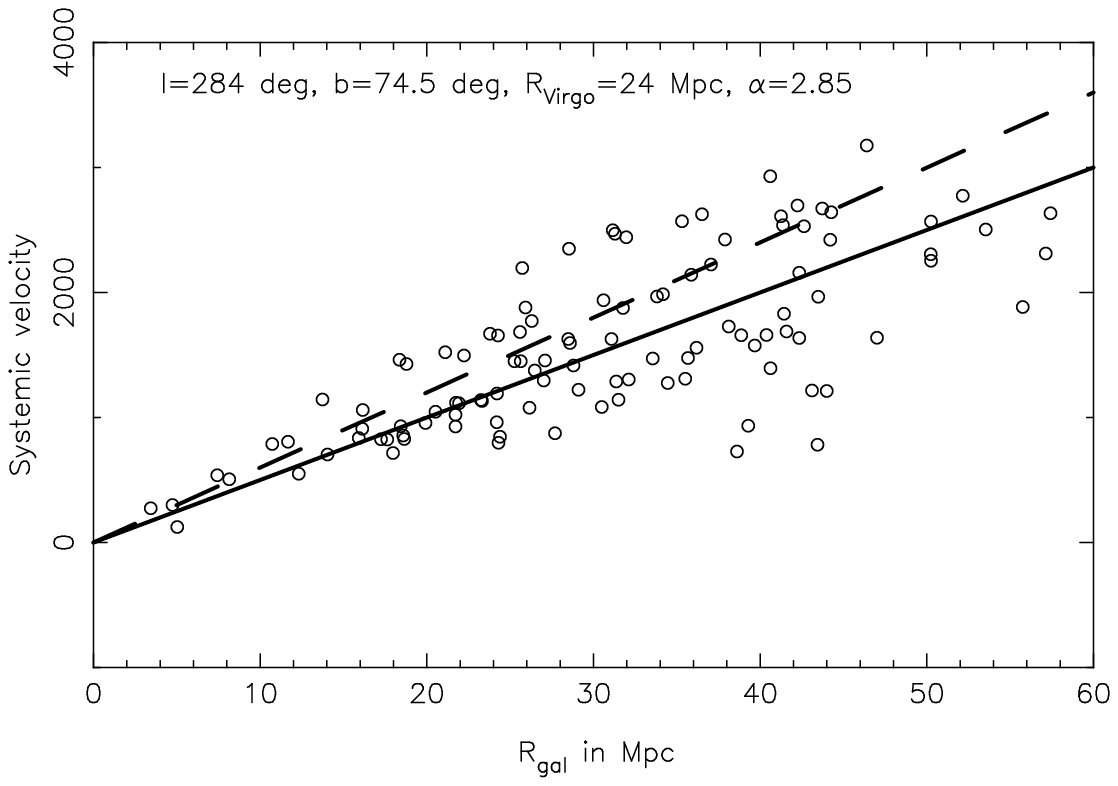}}
\caption{
The systemic velocity vs. distance for galaxies outside
the $\Theta=30\degr$ cone but still having
$V_\mathrm{obs}\le3\times V_\mathrm{Virgo}^\mathrm{cosm}\cos{\Theta}$
and $R_\mathrm{gal}\le60\mathrm{\ Mpc}$. The solid line is the
Hubble law predicted by
$H_0=50\mathrm{\ km\, s^{-1}\, Mpc^{-1}}$
and the dashed line the Hubble law predicted by
$H_0=60\mathrm{\ km\, s^{-1}\, Mpc^{-1}}$.
}
\label{F5}
\end{figure}

The observed systemic velocity vs. distance
$R_\mathrm{gal}$ diagrams are presented in Figs.~\ref{F1}-\ref{F5}.
In the first four figures galaxies belonging to a $\Theta<30\degr$
cone are shown for different angular intervals: galaxies having
$\Theta<10\degr$ are shown as black bullets, galaxies having
$10\degr\le\Theta<20\degr$ as grey bullets and galaxies having
$20\degr\le\Theta<30\degr$ as circles. The TB-curves are given
for the mean angular distance, $\langle\Theta\rangle$, for each
angular interval as thick black curve for $\langle\Theta\rangle=4.5\degr$,
as thick grey curve for $\langle\Theta\rangle=15.6\degr$ and as
thin black curve for $\langle\Theta\rangle=25.8\degr$. Comparison
between the data and the mean predictions were made for different presumed
distances to the centre of LSC:
$R_\mathrm{Virgo}=16\mathrm{\ Mpc}$ (Fig.~\ref{F1}),
$R_\mathrm{Virgo}=18\mathrm{\ Mpc}$ (Fig.~\ref{F2}),
$R_\mathrm{Virgo}=21\mathrm{\ Mpc}$ (Fig.~\ref{F3}) and
$R_\mathrm{Virgo}=24\mathrm{\ Mpc}$ (Fig.~\ref{F4}).
We remind that our model is formulated in terms of the relative
distance $d_\mathrm{gal}=R_\mathrm{gal}/R_\mathrm{Virgo}$. 
So the TB-curves show different
behaviour depending on the normalization.

The thick black
line in each figure corresponds to the Hubble law based on
$H_0=75\mathrm{\ km\, s^{-1}\, Mpc^{-1}}$,
$H_0=67\mathrm{\ km\, s^{-1}\, Mpc^{-1}}$,
$H_0=57\mathrm{\ km\, s^{-1}\, Mpc^{-1}}$ and
$H_0=50\mathrm{\ km\, s^{-1}\, Mpc^{-1}}$, respectively. The line
is drawn through the centre of LSC in order to emphasize our basic
assumption that the centre is at rest with respect to the cosmological
background. This also allows one to appreciate the infall of the Local
Group with an assumed velocity
$V_\mathrm{LG}^\mathrm{in}=220\mathrm{\ km\, s^{-1}}$.

%
%
%
%
%
%
%
%
Figs.~\ref{F1} and~\ref{F2} immediately reveal that the shorter
distances are not acceptable because the background galaxies fall
far below the expected curves. Correction for any residual
Malmquist bias would make situation even worse. Neither is 
$R_\mathrm{Virgo}=21\mathrm{\ Mpc}$, the distance found favourable
in Paper II, totally satisfying. Although the clump of galaxies at
$R_\mathrm{gal}\sim32\mathrm{\ Mpc}$ and 
$V_\mathrm{sys}\sim800\mathrm{\ km\, s^{-1}}$ 
in Fig.~\ref{F3} follow the
prediction as some other galaxies, the 
maximum of the velocity amplitude
is clearly behind the presumed centre.
%

This led us to test a longer distance to Virgo. The result
is shown in Fig.~\ref{F4}. It is rather remarkable that such
a distance gives better fit than the shorter ones. On the other
hand $R_\mathrm{Virgo}=24\mathrm{\ Mpc}$ together with the
adopted cosmological velocity bring about
$H_0=50\mathrm{\ km\, s^{-1}\, Mpc^{-1}}$. Such a small value
has for decades been advocated by Sandage and his collaborators
and is within the error bars of our determinations
(Theureau et al. \cite{Theureau97}; Ekholm et al.
\cite{Ekholm99b}) as well.
It is encouraging that galaxies outside the $30\degr$
cone follow well the Hubble law for this $H_0$. 
Virgo has only a weak
influence on them, and if the Malmquist    
bias is present these galaxies should
predict {\it larger} value for $H_0$. The dashed line in
Fig.~\ref{F5} is the Hubble law for 
$H_0=60\mathrm{\ km\, s^{-1}\, Mpc^{-1}}$. It is clearly an upper limit
thus giving us a lower limit for the
distance to Virgo: $R_\mathrm{Virgo}\ge20\mathrm{\ Mpc}$.  
%
%
%
%
\section{Have we found the true TB signature of Virgo?}
\begin{figure*}
\resizebox{\hsize}{!}{\includegraphics{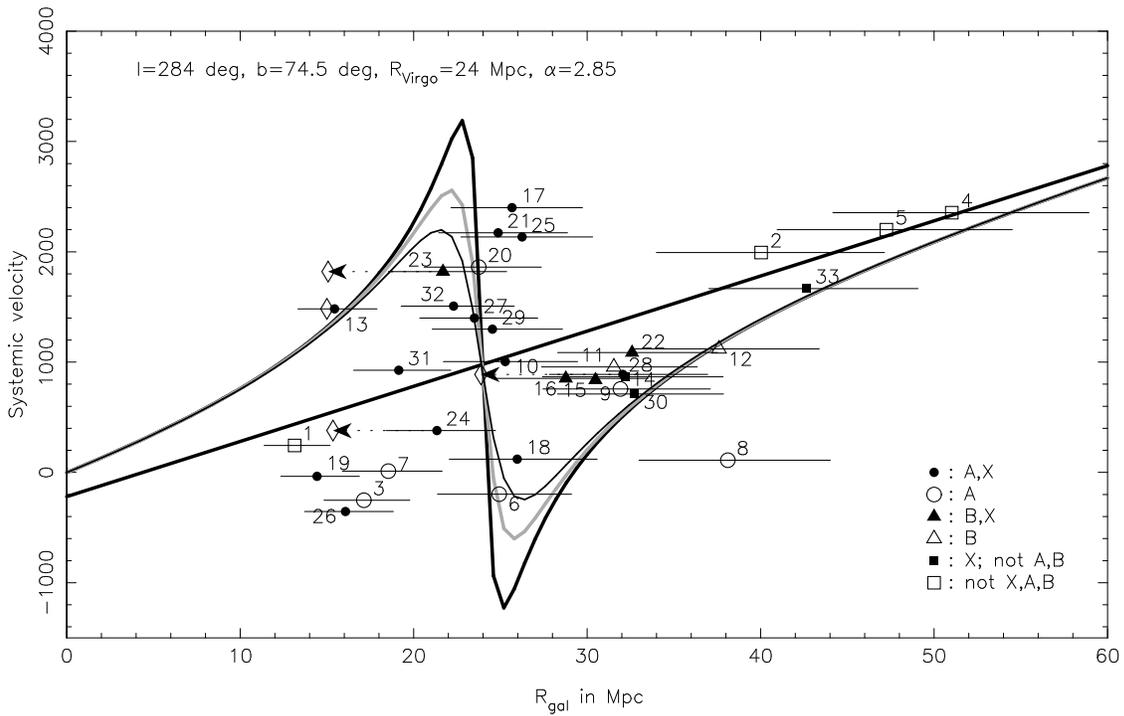}}
\caption{The systemic velocity vs. distance diagram for the
33 galaxies common in Table 1 in this paper
and Table~3 in Federspiel et al.
(\cite{Federspiel98}).}
\label{F6}
\end{figure*}

So far we have studied the $V_\mathrm{sys}$ vs. $R_\mathrm{gal}$
diagram in a simple way by moving the curves for the TB-solution
by choosing different normalizing distances to Virgo. The best
agreement with the maximum observed amplitude and the curves was
found at a rather large distance, namely 
$R_\mathrm{Virgo}=24\mathrm{\ Mpc}$. Such a long distance leads
one to ask whether we have actually found Virgo. We examine this
question by comparing our sample given in Table~1 with the sample
given by Federspiel et al. (\cite{Federspiel98}) from which they
found $R_\mathrm{Virgo}=20.7\mathrm{\ Mpc}$. We found 33
galaxies in common when requiring $\Theta<6\degr$. We present these
galaxies in Fig.~\ref{F6}. For an easy reference each galaxy
is assigned a number given also as a superscript after the name
in Table~1.

We give each galaxy a symbol following the classification of
Federspiel et al. (\cite{Federspiel98}). Following Binggeli et al.
(\cite{Binggeli93}) galaxies were divided into subgroup ``big A"
for galaxies close to M87 (`A') and into ``B" for galaxies within
$2\fdg4$ of M49 (`B'). They also examined whether a galaxy is
within the X-ray isophote $0.444\mathrm{\ counts\, s^{-1}\, arcmin^{-1}}$
based on ROSAT measurements of diffuse X-ray emission of hot gas
in the Virgo cluster (B\"ohringer et al. \cite{Bohringer94})
(`A,X', `B,X'). Galaxies belonging to subgroup A and 
within the X-ray contour are labelled in Fig.~\ref{F6} as bullets and
outside the contour with an open circle. Similarly, galaxies in
subgroup B are labelled with a filled or open triangle.
Federspiel et al. also listed galaxies within the X-ray contour
but not classified as members of A or B. The galaxies are marked
with a filled square. They also included in their Table~3 some
galaxies which fall outside A and B and the X-ray contour (we
label them with an open square).

We also give an error estimate for the TF-distance for each galaxy
calculated from the $1\sigma$ error in the distance modulus:
%
\begin{equation}
\label{E3}
\\ \sigma_\mu = \sqrt{\sigma_B^2+\sigma_{M_p}^2}.
\end{equation}
The error in the corrected total B-band magnitude is taken from
column (8) in our Table~1 and the intrinsic dispersion of the
absolute magnitude $M$ for each $p$, $\sigma_{M_p}$ is estimated
to be $0.3^\mathrm{mag}$.

The straight solid line is the Hubble law for
$H_0=50\mathrm{\ km\, s^{-1}\, Mpc^{-1}}$ shifted downwards
by $V_\mathrm{LG}^\mathrm{in}=220\mathrm{\ km\, s^{-1}}$
in order to make the line go through the centre at
$R_\mathrm{Virgo}=24\mathrm{\ Mpc}$ which is presumed to be at
rest with respect to the cosmological background. The TB-curves
are given for $\Theta=2\degr$ (thick black curve),
$\Theta=3\fdg5\degr$ (thick grey curve) and
$\Theta=5\degr$ (thin black curve), respectively.

To begin with, there are 22 galaxies (67 \%) 
which agree with the TB-solution
within $1\sigma$ in $\sigma_{\mu}$. Only four galaxies (1,3,19,26; 12 \%)
\footnote{Also 
\object{NGC 4216} (7) should probably be counted to this group,
because it differs by $\sim 2\sigma$ and clearly belongs to the same
substructure as the other four disagreeing galaxies. In other words,
15 \% of the sample does not agree with the model within
$2\sigma$.}  
do not agree with the model within $2\sigma$.
Though we have not reached the traditional 95 \% confidence level, 
the agreement is, at the
statistical level found, satisfying enough. 
Furthermore, we find in the range $R=24\pm2\mathrm{\ Mpc}$ nine
galaxies out of which seven were classified by Federspiel et al.
(\cite{Federspiel98}) as `A,X' galaxies hence presumably lying
in the very core of Virgo. 
The remaining two galaxies
are `A' galaxies. In the range $R=16\pm2\mathrm{\ Mpc}$ we find only
three `A,X' galaxies and one `A' galaxy. Federspiel et al.
(\cite{Federspiel98}) following Guhathakurta et al.
(\cite{Guhathakurta88}) listed five galaxies (17, 18, 19,
20 and 26 in Table~1) as HI-deficient. If these galaxies are
removed one finds four `A,X' and two `A' galaxies in the
range $R=24\pm2\mathrm{\ Mpc}$, and one `A,X' and one `A'
galaxy in the range $R=16\pm2\mathrm{\ Mpc}$. The numbers
are still clearly more favorable for a long distance to
Virgo. 

Four galaxies in this sample have also distances from the
extragalactic $PL$-relation (Lanoix \cite{Lanoix99};
Lanoix et al. \cite{Lanoix99a}, \cite{Lanoix99b}, \cite{Lanoix99c}).
These galaxies are NGC 4321 (13) with $R_{PL}=15.00\mathrm{\ Mpc}$,
NGC 4535 (23) with $R_{PL}=15.07\mathrm{\ Mpc}$, NGC 4548 (24) with
$R_{PL}=15.35\mathrm{\ Mpc}$ and NGC 4639 (28) with
$R_{PL}=23.88\mathrm{\ Mpc}$. These positions are shown as diamonds
in Fig.~\ref{F6}. The mean distance to Virgo using the
`A,X' galaxies 13, 21, 24, 27, 28, 29 and 32 (i.e. the 
HI-deficient galaxies excluded) with TF-distance moduli is
$\langle\mu\rangle=31.81$ or
$\langle R_\mathrm{Virgo}\rangle=22.98\mathrm{\ Mpc}$
and when using the $PL$-distance moduli available for the three
galaxies (13, 24 and 28) 
$\langle\mu\rangle=31.60$ or
$\langle R_\mathrm{Virgo}\rangle=20.93\mathrm{\ Mpc}$. The difference is
not large, and in both cases these Virgo core galaxies predict
a distance $R_\mathrm{Virgo}>20\mathrm{\ Mpc}$.
%
We find also some other interesting features in Fig.~\ref{F6}.
There are three galaxies (15, 16 and 22) which Federspiel et al.
(\cite{Federspiel98}) classified as `B,X' and two (11,12)
classified as `B'. Together they form a clearly distinguishable
substructure. It is the region D of Paper I, there
interpreted as a tight background concentration. The mean distance
for `B,X' galaxies is $\langle R\rangle=30.61\mathrm{\ Mpc}$
corresponding to $\mu=32.43$. This region is $0.53^\mathrm{mag}$
more distant than our presumed centre. We find this result satisfying
because Federspiel et al. (\cite{Federspiel98}) estimated that the
subgroup `B' (region D in Paper I) is, on average, about
$0.46^\mathrm{mag}$ farther distance than subgroup `A'. That our
sample brings about approximately the correct relative distance
between these subgroups lends additional credence to the distance
estimation made in the previous section.

The region B of Paper I described as an expanding component is
also conspicuously present in Fig.~\ref{F6} \footnote{
That such galaxies with negative velocity may be within a small
angular distance from the Virgo cluster and still be well in
the foreground was explained in Paper I as due to two things:
1) The expansion velocity must decrease away from the massive Virgo,
and 2) because of projection effects , the largest negative
velocities, belonging to galaxies at small distances from Virgo,
are seen close to the Virgo direction.}.
There is, however, no
clear trace of the region C of Paper I (galaxies of high velocities
but lying behind the centre; cf. Fig.~8 in Paper I)
unless NGC 4568 (25) actually lies at the same
distance as NGC 4567 ($R=34.28\mathrm{\ Mpc}$). It should be
remembered that NGC 4567/8 is classified as an interacting pair.
There are, however, in Fig.~\ref{F4} many galaxies at larger angular
distances around NGC 4567. It is possible that they form the
region C. In Paper I region C was divided into two subregions,
C1 and C2. C1 was interpreted as the symmetrical counterpart
to the region B (these galaxies behind Virgo are expanding away
from it) and C2 was considered as a background contamination.
Galaxies in region A (galaxies with high velocities lying in
front of the centre) were proposed in Paper I to be presently
falling into Virgo. 
As regards regions A and C1 it is now easy
to understand that they are not separate regions but reflect the
behaviour of the TB-curve: A is on the rising part and C1 on the
declining part of the curve in front of the structure.

We conclude that from the expected distance-velocity pattern 
we have accumulated quite convincing evidence for
a claim that the distance to the Virgo cluster is 
$R_\mathrm{Virgo}=20$ -- $24\mathrm{\ Mpc}$ or in terms of the
distance modulus $\mu=31.51$ -- $31.90$. $\Delta\mu=0.39$ is within
$1\sigma$ uncertainty of our TF-sample. 
%
%
%
%
\section{The velocity field as seen from the centre of LSC}
In the first part of this paper we have approached 
the problem of the dynamical behaviour
of LSC in a more or less qualitative manner. We now
proceed to present the results in a physically more relevant
manner. 
The main
difficulty in the presentation used e.g. in Figs.~\ref{F1}-~\ref{F4}
is that the systemic velocity depends not only on the distance
from LG but also on the angular distance from the centre.
Basically, for each galaxy there is a unique ``S-curve"
depending on $\Theta$. 

Formally, the $\Theta$-dependence is removed if the velocity-distance
law is examined from the origin of the metric instead of from LG,
as was done in Sect. 4.5 of Paper I. The velocity as seen from
Virgo for a galaxy is solved from: 
%
\begin{equation}
\label{E4}
\\ v(d_\mathrm{c}) = \pm
\frac{V_\mathrm{obs}(d_\mathrm{gal})-V_\mathrm{Virgo}^\mathrm{obs}\cos\Theta}
{\sqrt{1-\sin^2\Theta/d_\mathrm{c}^2}}.
\end{equation}
The relative distance from the centre 
$d_\mathrm{c} = R_\mathrm{c}/R_\mathrm{Virgo}$, where
$R_\mathrm{c}$ is the distance between the galaxy considered
and the centre of Virgo, 
is solved from Eq.~14 of Paper II and the
sign is $(-)$ for $d_\mathrm{gal}<\cos\Theta$ and $(+)$
otherwise. 

There are, however, some difficulties involved.
We are aware that the calculation of the virgocentric velocity 
is hampered by some sources of error. Suppose that the
cosmological fluid has a perfect radial symmetry about the origin 
of the TB-metric. Also, the fluid elements do not interact with
each other, i.e. each element obeys {\it exactly} the equations of
motion of the TB-model. 
It follows that the measured line-of-sight
velocity is a genuine projection of the element's velocity with
respect to the origin. It is presumed that the observer has
made the adequate corrections for the motions induced by his
immediate surroundings (e.g. Sun's motion with respect to the Galaxy,
Galaxy's motion with respect to the LG).

Now, in practice,
$V_\mathrm{obs}$ is bound to contain also other
components than simply the TB-velocity. We may also have  mass shells 
which have travelled through the origin and are presently expanding
near it instead being falling in. Such a shell has experienced
strong pressures (in fact, a singularity has formed to the origin)
i.e. there is no causal connection to the rest of
the TB-solution. Also, shells may have crossed. 
Again singularity has formed and the TB-solution fails (recall
that TB-model describes a {\it pressure-free} cosmological fluid).  
Incorrect distance $R_\mathrm{gal}$ (and the scaling
length $R_\mathrm{Virgo}$) will cause an error in
$v(d_\mathrm{c})$ even when $V_\mathrm{obs}$ could be
considered as a genuine projection of $v(d_\mathrm{c})_\mathrm{TB}$.
%
\begin{figure*}
\resizebox{12cm}{!}{\includegraphics{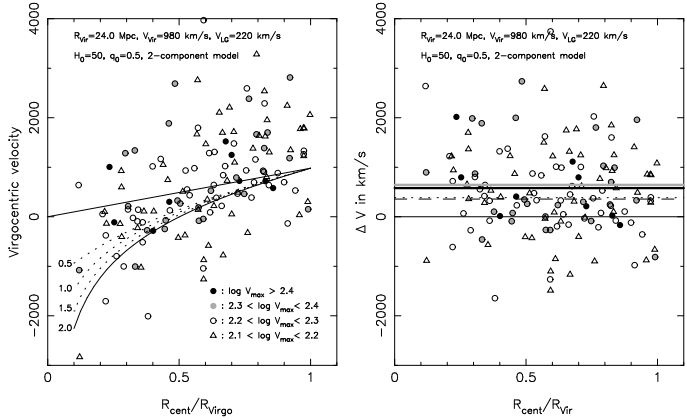}}
\hfill
\parbox[b]{55mm}{
\caption{Left panel: The virgocentric velocity as a 
function of TF-distance
from the centre for $R_\mathrm{Virgo}=24\mathrm{\ Mpc}$.
The solid line is the Hubble law one would
see from the centre and the curves are the TB-predictions for the
two-component model (for details cf. text).
Right panel: comparison between the calculated and predicted 
($\beta=2.0$, $\alpha=2.5$) virgocentric velocities.
}
\label{F7}}
\end{figure*}
\subsection{The two-component mass model}
So far we have used a rather simple density model. From hereon
we use the ``two-component" model of Paper I. In this model
one assumes that mass within $\Theta=6\degr$ at Virgo distance
($d_\mathrm{virial}=0.105$) is proportional to the Virgo virial
mass and that outside this region
the mass is evaluated from the simple density law
(Eq.~9 in Paper II):
%
\begin{equation}
\label{E5}
\\ M(d_\mathrm{c}) = 
M(d_\mathrm{c})_\alpha-M(d_\mathrm{virial})_\alpha+\beta M_\mathrm{virial}.
\end{equation}
The important quantity is the parameter $A(R,T_0)$ (Eq.~6 in Paper II).
Following Ekholm (\cite{Ekholm96}) 
we now proceed to express it in terms of the
relative distance ``measured" from the origin of the metric 
$d\equiv d_\mathrm{c}$ 
and the deceleration parameter $q_0$.
In terms of $d$ it reads:
%
\begin{equation}
\label{E6}
\\ A(d,T_0) = \sqrt{\frac{GM(d)}{d^3R_\mathrm{Virgo}^3}}\times T_0.
\end{equation}
Because (cf. Eq.~9 in Paper II)
%
\begin{equation}
\label{E7}
\\ M(d)_\alpha=\frac{q_0 H_0^2}{G}d^3 R_\mathrm{Virgo}^3[1+k'\ d^{-\alpha}],
\end{equation}
we find
%
\begin{eqnarray}
\label{E8}
A(d,T_0) & = & H_0 T_0\sqrt{q_0}\,[ 1+k'd^{-\alpha}- 
(d_\mathrm{virial}/d)^3 (1+k'd_\mathrm{virial}^{-\alpha}) \nonumber \\
& &+(\beta GM_\mathrm{virial})/(d^3 R_\mathrm{Virgo}^3 H_0^2)]^{1/2}. 
\end{eqnarray}
Now, using $M_\mathrm{virial}=7.5\times10^{14}M_{\sun}
R_\mathrm{Virgo}/16.8\mathrm{\ Mpc}$, $H_0 T_0=C(q_0)$ (e.g. the function
$C(q_0)=2/3$ for $q_0=0.5$)
and $H_0 R_\mathrm{Virgo}=V_\mathrm{Virgo, cosm}$, Eq.~\ref{E8} 
takes its final form
%
\begin{eqnarray}
\label{E9}
A(d,q_0) & = & C(q_0)\sqrt{q_0}\,[ 1+k'd^{-\alpha}- 
(d_\mathrm{virial}/d)^3 (1+k'd_\mathrm{virial}^{-\alpha}) \nonumber \\
& &+(\beta\times \mathrm{cst})
/(d^3 V_\mathrm{Virgo, cosm}^2)]^{1/2}, 
\end{eqnarray}
where $\mathrm{cst}
=7.5\times10^{14}M_{\sun}G/16.8\mathrm{\ Mpc}
=1.92\times10^5\mathrm{\ km^2\,s^{-2}}$. 
\subsection{$v(d_\mathrm{c})$ vs. $d_\mathrm{c}$ diagram for
$R_\mathrm{Virgo}=24\mathrm{\ Mpc}$}
We show the virgocentric diagram for 
$R_\mathrm{Virgo}=24\mathrm{\ Mpc}$ in
the left panel of Fig.~\ref{F7}. 
The galaxies are now selected in the following manner.
From the initial sample we take galaxies having
$0.105<d_\mathrm{c}\le1.0$ but make no restriction on $\Theta$. 
In this way we get a symmetric sample around the centre.
Because the angular dependence
is no longer relevant, we show the data for different ranges of
$\log V_\mathrm{max}$: black bullets are for
$\log V_\mathrm{max}\ge 2.4$, grey bullets for
$\log V_\mathrm{max}\in[2.3, 2.4[$, circles for
$\log V_\mathrm{max}\in[2.2,2.3[$ and triangles for
$\log V_\mathrm{max}\in[2.1,2.2[$. 
The straight line is Hubble law
as seen from the centre and the curves 
(predicted velocity $v'(d_\mathrm{c})$ vs. $d_\mathrm{c}$)
correspond to different
solutions to the two-component model. We have assumed $\alpha=2.5$
and solved the TB-equations with Eq.~\ref{E9} for
$\beta=0.5$, $1.0$, $1.5$ and $2.0$ yielding mass excesses
$k'=0.701$, $0.504$, $0.307$ and $0.109$, respectively.

Because the gradient of the $v'(d_\mathrm{c})$-curve gets quite
steep as $d_\mathrm{c}\rightarrow 0$, it is easier to study
the difference between calculated and predicted velocities 
%
\begin{equation} 
\label{E10}
\\ \Delta v(d_\mathrm{c}) = v(d_\mathrm{c})-
v'(d_\mathrm{c})
\end{equation}
as a function of $d_\mathrm{c}$. This is shown in the right panel
of Fig.~\ref{F7}. The model values $v'(d_\mathrm{c})$ were based
on $\beta=2.0$. In this panel we also show the mean $\Delta v$
for each $\log V_\mathrm{max}$ range.   
For $\log V_\mathrm{max}\ge 2.4$ 
($N=9$, $\Delta v=579\mathrm{\ km\, s^{-1}}$) 
it is given as a black thick line,
for $\log V_\mathrm{max}\in[2.3,2.4[$
($N=26$, $\Delta v=646\mathrm{\ km\, s^{-1}}$) 
as a grey thick line,
for $\log V_\mathrm{max}\in[2.2,2.3[$ 
($N=76$, $\Delta v=359\mathrm{\ km\, s^{-1}}$) 
as a dashed line,
and for $\log V_\mathrm{max}\in[2.1,2.2[$ 
($N=55$, $\Delta v=385\mathrm{\ km\, s^{-1}}$) 
as a dotted line. We note that our sample is
clearly divided into two subgroups by $\log V_\mathrm{max}=2.3$. The
slower rotators show a better fit to our chosen model. In general,
galaxies in this sample have on average higher velocities than the
model predicts, possibly due to some residual Malmquist bias
(cf. also Figs.~5 -- 6 of Paper I). 
It is, however, clear that the overall TB-pattern
is seen in the left panel of Fig.~\ref{F7} as a general decrease
in $v(d)_\mathrm{c}$ when one approaches the centre.
%
\begin{figure*}
\resizebox{12cm}{!}{\includegraphics{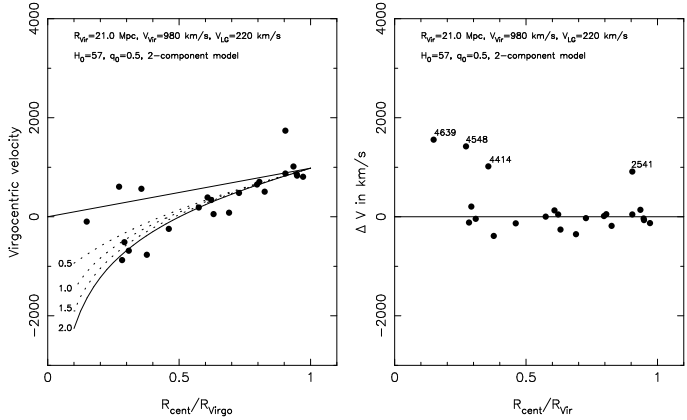}}
\hfill
\parbox[b]{55mm}{
\caption{Left panel: The virgocentric velocity vs. distance for
the 23 galaxies with $PL$-distances. The relative distances are
based on $R_\mathrm{Virgo}=21\mathrm{\ Mpc}$. Right panel:
comparison between calculated and predicted velocities for
$\beta=2.0$, $\alpha=2.5$. 
}
\label{F8}}
\end{figure*}
%
%
%
\begin{figure*}
\resizebox{12cm}{!}{\includegraphics{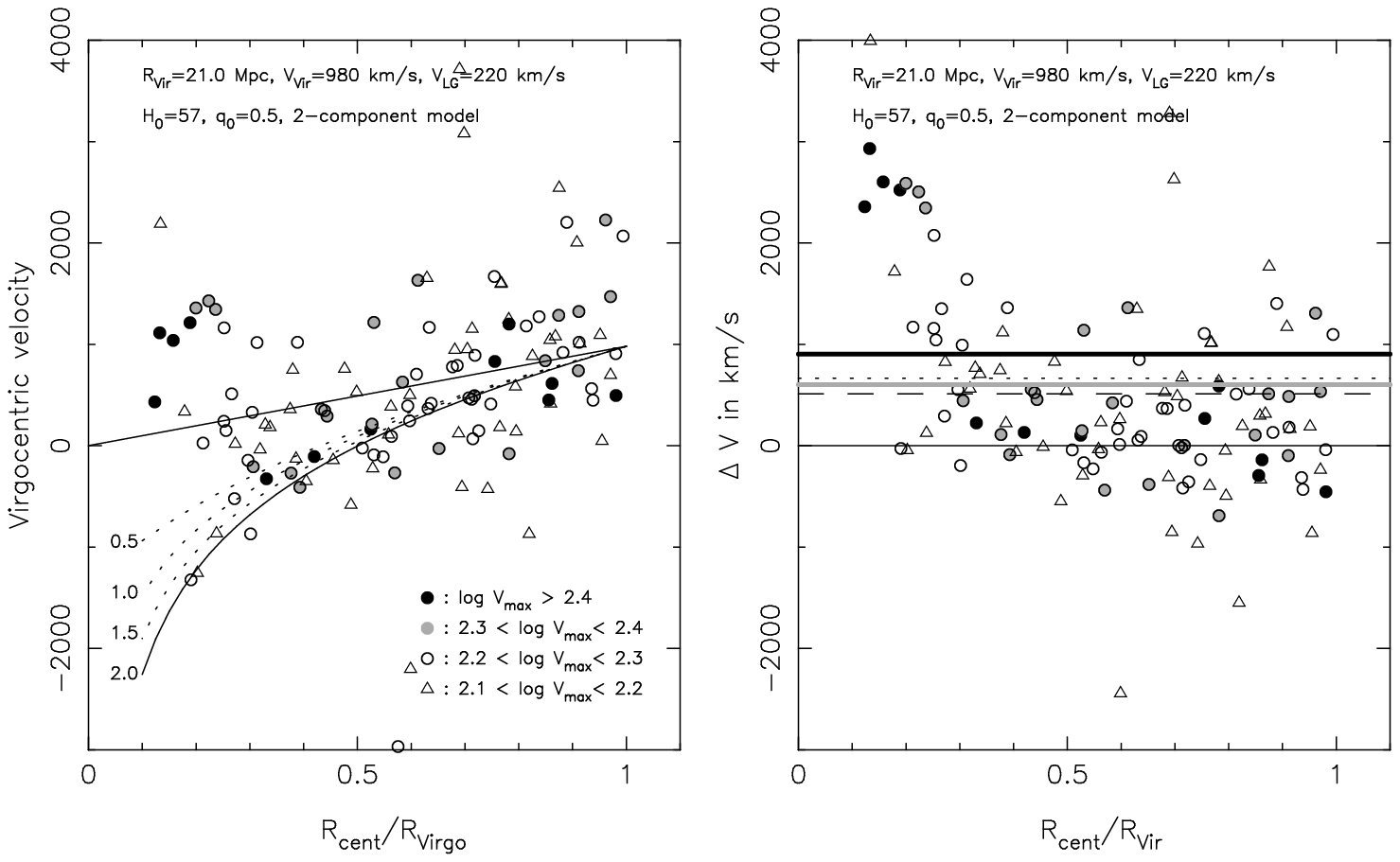}}
\hfill
\parbox[b]{55mm}{
\caption{As Fig.~\ref{F7}, but now the distance to Virgo is
$R_\mathrm{Virgo}=21\mathrm{\ Mpc}$.
}
\label{F9}}
\end{figure*}
\subsection{Evidence from galaxies with $PL$-distances}
How do the galaxies with $PL$-distances behave in this
virgocentric representation? When selected in a similar fashion
as above we find 23 galaxies shown in Fig.~\ref{F8}.
We saw that $R_\mathrm{Virgo}=24\mathrm{\ Mpc}$ was
a rather high value for them
but now $R_\mathrm{Virgo}=21\mathrm{\ Mpc}$
together with $\alpha=2.5$ and $\beta=2.0$ brings about a
remarkable accordance. This is particularly important
in the light of the complications mentioned in the
introduction to this section. It seems that at least
when using high quality distances such as $PL$-distances
those difficulties do not hamper the diagrams
significantly. 
When this result is compared with the
findings of Paper II, the distance estimate given there seems
to be more and more acceptable. 

There are four galaxies which show anomalous behaviour.
\object{NGC 2541} is a distant galaxy as seen from Virgo
($R_\mathrm{gal}=11.59\mathrm{\ Mpc}$, $\Theta=63.8\degr$,
$V_\mathrm{obs}=645\mathrm{\ km\, s^{-1}}$) 
and is also close to the tangential point
where small errors
in distance cause large projection errors in velocity.
We tested how much one needs to move this galaxy in order to
find the correct predicted velocity. 
At $R_\mathrm{gal}=13.93\mathrm{\ Mpc}$, 
$V_\mathrm{pred}=645.1\mathrm{\ km\, s^{-1}}$ and
$v(d_\mathrm{c})=884.7\mathrm{\ km\, s^{-1}}$ with
$\Delta v=-0.5\mathrm{\ km\, s^{-1}}$. Note also
that even a shift of $1\mathrm{\ Mpc}$ to 
$R_\mathrm{gal}=12.59\mathrm{\ Mpc}$ will yield
$\Delta v=360.0\mathrm{\ km\, s^{-1}}$, which is
quite acceptable. When \object{NGC 4639}
($R_\mathrm{gal}=23.88\mathrm{\ Mpc}$, $\Theta=3.0\degr$,
$V_\mathrm{obs}=888\mathrm{\ km\, s^{-1}}$) is moved to
$R_\mathrm{gal}=21.0\mathrm{\ Mpc}$, one finds
$V_\mathrm{pred}=886.8\mathrm{\ km\, s^{-1}}$ and
$v(d_\mathrm{c})=-3419.5\mathrm{\ km\, s^{-1}}$ with
$\Delta v=45.6\mathrm{\ km\, s^{-1}}$. What is
interesting in this shift is that in Paper II most of
the galaxies tended to support 
$R_\mathrm{Virgo}=21\mathrm{\ Mpc}$ except this galaxy
and \object{NGC 4548}. Now \object{NGC 4639} fits
perfectly. Recently Gibson et al. (\cite{Gibson99})
reanalyzed some old HST measurements finding for
\object{NGC 4639}: $\mu=31.564$ or 
$R_\mathrm{gal}=20.55\mathrm{\ Mpc}$. As regards the two
other discordant galaxies (\object{NGC 4414} and
\object{NGC 4548}) the shift to remove the discrepancy
would be too large to be reasonable. At this point we cannot 
explain their behaviour except by assuming that they are
region B galaxies of Paper I (cf. below). 

Also, when galaxies with TF-distances were selected according
to this normalizing distance we find better concordance
with the model than for $R_\mathrm{Virgo}=24\mathrm{\ Mpc}$
(cf. Fig.~\ref{F9}). Note also that now only the fastest
rotators differ from the rest of the sample:  
for $\log V_\mathrm{max}\ge 2.4$: 
$N=12$, $\Delta v=904\mathrm{\ km\, s^{-1}}$ 
(black thick line),
for $\log V_\mathrm{max}[2.3,2.4[$:
$N=23$, $\Delta v=602\mathrm{\ km\, s^{-1}}$ 
(grey thick line),
for $\log V_\mathrm{max}\in[2.2,2.3[$: 
$N=65$, $\Delta v=512\mathrm{\ km\, s^{-1}}$
(dashed line) 
and for $\log V_\mathrm{max}\in[2.1,2.2[$: 
$N=49$, $\Delta v=665\mathrm{\ km\, s^{-1}}$
(dotted line).
At relatively large distances from the centre the points in
the right panel of Fig.~\ref{F9} follow on average well a
horizontal trend. As one approaches the centre one sees
how the velocity difference $\Delta v$ gets larger and larger.
This systematic increase explains why the mean values are so
high. Note that also the Cepheid galaxies \object{NGC 4414} and
\object{NGC 4548} (and \object{NGC 4639} if one accepts the
larger distance) show a similar
increasing tendency towards the centre.
 
Because the inward growth of $\Delta v$ 
appears for both distance indicators one suspects that
this behaviour is a real physical phenomenon
(we cannot explain it in terms of a large scatter in the 
TF-relation). Neither can we explain it by a bad choice of
model parameters: the effect is much stronger than the variations
between different models. 
A natural explanation is an
expanding component (referred to above as region B): galaxies with
very high $\Delta v$ are on mass shells which have fallen through
the origin in past and have re-emerged as a ``second generation"
of TB-shells. The very quick decay of the positive velocity
residuals supports this picture. The mass of the Virgo
cluster is expected to slow down these galaxies quite 
fast (Sect.~6 in Paper I), so the effect appears at small
$d_\mathrm{c}$.
%
%
%
%
\section{Discussion}
We found using the two-component mass model 
(Eq.~\ref{E5}) and the high
quality $PL$-distances (Fig.~\ref{F8}) an acceptable fit
with parameters $\alpha=2.5$ and $\beta\approx2.0$.
Our larger TF-sample did not disagree with this model
though the scatter for these galaxies is rather large. 
$\beta$ gives the Virgo cluster mass estimate in terms of the
virial mass given by Tully \& Shaya (\cite{Tully84}). With
a distance $R_\mathrm{Virgo}=21\mathrm{\ Mpc}$ it
is $M_\mathrm{TS}=9.375\times10^{14}M_{\sun}$. By
allowing some tolerance ($\beta=1.5$ -- $2.0$)
we get an estimate:
%
\begin{equation}
\label{E11}
\\ M_\mathrm{Virgo}=(1.4\mathrm{\ -\ }1.875)\times10^{15}M_{\sun}
\end{equation}
\subsection{The Virgo cluster mass, $q_0$, and behaviour of $M/L$}
We have confirmed the large value of the mass-luminosity ratio
for the Virgo cluster (Tully \& Shaya \cite{Tully84}; Paper I):
%
\begin{equation}
\label{E12} 
\\ (M/L)_\mathrm{Virgo}\approx 440\beta\times
(16.8\mathrm{\ Mpc}/R_\mathrm{Virgo}).
\end{equation}
With $\beta = 1$ -- $2$ and 
$R_\mathrm{Virgo} = 21\mathrm{\ Mpc}$, 
$(M/L)_\mathrm{Virgo}$ ranges from 350 to 700.  
Note that
some calculations of Paper I for different $q_0$ (e.g. Table 2), 
which were based on
$H_0 = 70\mathrm{\ km\, s^{-1}\, Mpc^{-1}}$ and 
$R_\mathrm{Virgo} = 16.5\mathrm{\ Mpc}$,
remain valid when 
$H_0 = 55\mathrm{\ km\, s^{-1}\, Mpc^{-1}}$ and 
$R_\mathrm{Virgo} = (70/55)\times 16.5\mathrm{\ Mpc} = 21\mathrm{\ Mpc}$.
For example, Fig.~7 of Paper I shows that if 
$(M/L)_\mathrm{Virgo}$ applies everywhere,
rather high values of $q_0$ ($ > 0.1$ -- $0.2$) 
are favoured. A very small
$q_0$, say 0.01, would require that $M/L$ outside of Virgo is several times
smaller than in Virgo, i.e. the density of dark matter drops much more
quickly than the density of luminous matter.

This happens also -- though less rapidly -- with $q_0 = 0.5$ 
used in this paper. This is seen from 
%
\begin{equation}
\label{E13} 
\\ \frac{(M/L)_\mathrm{sur}}{(M/L)_\mathrm{Virgo}}= 
\frac{M_\mathrm{sur}}{M_\mathrm{Virgo}}\times
\frac{L_\mathrm{Virgo}}{L_\mathrm{sur}}.
\end{equation}
The surroundings is defined as $d_\mathrm{c}\in\ ]0.105,1[$.
The luminosity ratio is
$L_\mathrm{Virgo}/L_\mathrm{sur} \approx 1/4$ 
(Tully \cite{Tully82}). The mass ratio is
calculated using the two-component mass model (Eq.~\ref{E5})
with the help of Eq.~18 of paper II. For $M_\mathrm{Virgo}=2$ 
the parameters needed are $k'=0.109$, 
$\alpha=2.5$, $q_0=0.5$ and $h_0=0.57$, which yield 
$M(d_\mathrm{c}=1)_{\alpha}=4.155$ and
$M(d_\mathrm{c}=0.105)_{\alpha}=0.137$.
We find
$M_\mathrm{sur}=M(d_\mathrm{c}=1)_{\alpha}-
M(d_\mathrm{c}=0.105)_{\alpha}=4.018\approx4$.
Both masses are
given in units
of the Virgo virial mass. The mass-luminosity ratio becomes 
$(M/L)_\mathrm{sur}/(M/L)_\mathrm{Virgo}\approx 0.5$.
When $M_\mathrm{Virgo}=1$
$(M/L)_\mathrm{sur}/(M/L)_\mathrm{Virgo}\approx 1.25$
and when
$M_\mathrm{Virgo}=1.5$ 
$(M/L)_\mathrm{sur}/(M/L)_\mathrm{Virgo}\approx 0.75$.\footnote{The
total mass within $d_\mathrm{c}=1$ is 6.018 for $\beta=2$,
6.019 for $\beta=1$ and 6.020 for $\beta=1.5$. The Model 1 of
Paper II ($k'=0.606$ and $\alpha=2.85$) gives 6.017 as the total
mass. Thus our computational scheme works correctly because the
total mass should not depend on how we distribute the matter
within our mass shell.} This means that with a Virgo mass
slightly larger than the virial mass there is a case where
the mass-luminosity ratio is constant in and outside Virgo.

How would luminous matter distribute itself? Consider the following
simple exercise. Suppose the luminous matter follows
a power law 
$\rho_\mathrm{lum}(r) \propto r^{-\alpha_\mathrm{lum}}$ and that
the mass ratio is:
%
\begin{equation}
\label{E14}
\\ \frac{\int_{0.105}^{1} r^{2-\alpha} dr}
{\int_{0}^{0.105} r^{2-\alpha} dr}=
\frac{L_\mathrm{sur}}{L_\mathrm{Virgo}}.
\end{equation}
With the luminosity ratio
given above one derives for
the galaxies $\alpha_\mathrm{lum} \approx 2.3$, 
indeed smaller than our preferred value
of 2.5. Is such a steep value at all reasonable in the light of
theoretical work on structure formation?
\subsection{Comparison with the universal density profile}
Tittley \& Couchman (\cite{Tittley99}) discussed recently
the hierarchical clustering, the universal density profile, and
the mass-temperature scaling law of galaxy clusters. Using
simulated clusters they studied the dark matter density profile
in a Einstein-deSitter universe with $\Omega_{DM}=0.9$,
$\Omega_\mathrm{gas}=0.1$ and $\Lambda=0$. They assumed
$H_0=65\mathrm{\ km\, s^{-1}\, Mpc^{-1}}$. Different profiles
fitted their simulated data equally well. It is their 
discontinuous form in the first derivative which interests us:
%
\begin{equation}
\label{E15}
\frac{\rho(r)}{\rho_c}=\left\{
\begin{array}{ll}
\delta_{\gamma'}r^{-\gamma'}, & r<r_s \\
\delta_{\gamma}r^{-\gamma}, & r>r_s
\end{array}
\right.
\end{equation}
They connect the overdensities as
%
\begin{equation}
\label{E16}
\\ \delta_{\gamma} = \frac{r_s^{\gamma}}{r_s^{\gamma'}}\delta_{\gamma'}.
\end{equation}
Because the characteristic length $r_s<R_{200}$, where $R_{200}$
is the radius where the density contrast equals 200, the near field
governed by $\gamma$ is not important to us. With $\alpha=2.5$ and
$\beta=2.0$ in our model the mass excess $k'=0.109$. This translates
into $k=(3-\alpha)\times k'R_\mathrm{Virgo}^{\alpha}/3=36.71$
in the density law of Paper II: 
$\delta(r) = \rho(r) /\rho_0=1+kr^{-\alpha}$.
$\rho_0$ is the background density equal to the
critical density $\rho_c$ when $q_0=0.5$.
At the defined boundary of the Virgo cluster
($d=0.105$ or $r=2.205\mathrm{\ Mpc}$) we have a density excess  
$\delta=5$. For $\beta=1.5$, $k=103.4$ and $\delta=14.32$, and
for $\beta=1.0$, $k=168.4$ and $\delta=23.33$. 
Also, because $1+kr^{-\alpha}\rightarrow kr^{-\alpha}$ as
$r\rightarrow r_s$ comparison between our $\alpha$ and the
$\gamma$ of Tittley \& Couchman is acceptable.
For hierarchical clustering they find $\gamma=2.7$ and
for the non-hierarchical case $\gamma=2.4$. The density profile fitting
dynamical behaviour of the galaxies with $PL$-distances is
within these limits. Our mass estimate tends to be closer to
the maximum values Tittley \& Couchman give in their Table~3. 
%
%
%
%
\section{Summary and conclusions}
In this third paper of our series we have extended the discussion
of Ekholm et al. (\cite{Ekholm99a}; Paper II) to the background of
Virgo cluster by selecting galaxies with 
as good distances as possible from the direct B-band magnitude
Tully-Fisher (TF) relation. 
In the following list we summarize our main results:
\begin{enumerate}
\item Although having a rather large scatter the TF-galaxies
reveal the expected Tolman-Bondi (TB) pattern well. We compared our
data with TB-solutions for different distances to the Virgo
cluster. It turned out that when $R_\mathrm{Virgo}<20\mathrm{\ Mpc}$
the background galaxies fell clearly below the predicted curves. Hence the
data does not support such distance scale (cf. Figs.~\ref{F1}
and~\ref{F2}).
\item When we examined the Hubble diagram for galaxies outside
the Virgo $\Theta=30\degr$ cone (Fig.~\ref{F5}) we noticed
that $H_0=60\mathrm{\ km\, s^{-1}\, Mpc^{-1}}$ is a clear
upper limit for these galaxies. Together with our preferred
cosmological velocity of Virgo ($1200\mathrm{\ km\, s^{-1}}$)
we concluded that $R_\mathrm{Virgo}=20\mathrm{\ Mpc}$ is a
lower limit.
\item In both cases any residual Malmquist bias would move the
sample galaxies further away and thus make the short distances
even less believable.
\item We compared our sample galaxies with $\Theta<6\degr$ with the Table
3 of Federspiel et al. (\cite{Federspiel98}) and found 33 galaxies
in common. We established a plausible case for 
$R_\mathrm{Virgo}=24\mathrm{\ Mpc}$ corresponding to
$H_0=50\mathrm{\ km\, s^{-1}\, Mpc^{-1}}$ (cf. Fig.~\ref{F6}). 
The difference between
$R_\mathrm{Virgo}=20\mathrm{\ Mpc}$ and 
$R_\mathrm{Virgo}=24\mathrm{\ Mpc}$ is -- in terms of the distance
moduli -- only $\Delta\mu=0.39$, which is within the $1\sigma$
scatter of the TF-relation. Due to this scatter it is not possible
to resolve the distance to Virgo with higher accuracy. Hence we
claim that $R_\mathrm{Virgo}=20$ -- $24\mathrm{\ Mpc}$.
\item Some of the kinematical features identified in Paper I were
revealed also here, in particular the concentration of galaxies in
front with very low velocities (interpreted as an expanding component;
region B in Paper I)
and the tight background concentration (region D in Paper I).
The symmetric counterpart of region B (region C1) may actually
be part of the primary TB-pattern.
\item The need
for a better distance indicator (e.g. the I-band TF-relation) is
imminent. As seen e.g. from Fig.\ref{F9}, the scatter in the
B-band TF-relation is disturbingly large. It is also necessary
to re-examine the calibration of the TF-relation with the new,
and better, $PL$-distances. It seems that the $PL$-distances
and the TF-distances from Theureau et al. (\cite{Theureau97})
are not completely consistent. The former tend to be somewhat
smaller. This is also seen from Figs.~\ref{F6} and~\ref{F8}.
TF-distances support $R_\mathrm{Virgo}= 24 \mathrm{\ Mpc}$ 
and $PL$-distances $R_\mathrm{Virgo}= 21 \mathrm{\ Mpc}$.
It is, however, worth reminding that our dynamical conclusions
are insensitive to the actual distance scale.
\item When we examined the Hubble diagram as it would be seen
from the origin of the TB-metric, galaxies with distances from
the extragalactic $PL$-relation fitted best to a solution with
$R_\mathrm{Virgo}=21\mathrm{\ Mpc}$ in concordance with Paper II
and with Federspiel et al. (\cite{Federspiel98}). We are, however,
not yet confident enough to assign any error bars to this value.
\item For $R_\mathrm{Virgo}=21\mathrm{\ Mpc}$ the region D follows
well the TB-pattern (cf. Fig.~\ref{F3}) lending some additional
credence to this distance. We quite clearly identified this
background feature as the subgroup ``B" of Federspiel et al.
(\cite{Federspiel98}).
\item These high quality galaxies also clearly follow the expected
velocity-distance behaviour in the virgocentric frame with much
smaller scatter than for galaxies in Paper I
or for the TF-galaxies used in this paper. The zero-velocity
surface was detected at $d_\mathrm{c}\approx 0.5$.
\item As in Teerikorpi et al. (\cite{Teerikorpi92}; Paper I), the
amplitude of the TB-pattern requires that the Virgo cluster mass must
be at least its standard virial mass (Tully \& Shaya \cite{Tully84})
or more. Our best estimate is
$M_\mathrm{Virgo}=(1.5$ -- $2)\times M_\mathrm{virial}$, where
$M_\mathrm{virial}=9.375\times10^{14}M_{\sun}$ for
$R_\mathrm{Virgo}=21\mathrm{\ Mpc}$.
\item Our results indicate that the density distribution of
luminous matter is shallower than that of the total gravitating
matter. The preferred exponent in the density power law,
$\alpha\approx2.5$, agrees with the theoretical work on the
universal density profile of dark matter clustering
(Tittley \& Couchman \cite{Tittley99}) in the
Einstein-deSitter universe.
\end{enumerate}
%
%
%
%
\begin{acknowledgements}{
This work has been partly supported by the Academy of Finland
(project 45087: ``Galaxy Streams and Structures in the nearby
Universe" and project ``Cosmology in the Local Galaxy Universe").
We have made use of the Lyon-Meudon Extragalactic Database
LEDA and the Extragalactic Cepheid Database. We would like to
thank the referee for useful comments.}
\end{acknowledgements}
%
%
%

%
%

\begin{thebibliography}{}
%
\bibitem[1993]{Binggeli93}
Binggeli,~B., Popescu,~C.~C., Tammann,~G.~A., 1993, A\&AS 98, 275
%
\bibitem[1994]{Bohringer94}
B\"ohringer,~H., Briel,~U.~G., Schwartz,~R.~A. et al., 1994, Nat 368, 828
%
\bibitem[1947]{Bondi47}
Bondi,~H., 1947, MNRAS, 107, 410
%
\bibitem[1996]{Ekholm96}
Ekholm,~T., 1996
A\&A  308, 7
%
\bibitem[1994]{Ekholm94}
Ekholm,~T., Teerikorpi,~P., 1994,
A\&A 284, 369
%
\bibitem[1999a]{Ekholm99a}
Ekholm,~T., Lanoix,~P., Teerikorpi,~P. et al., 1999a, 
A\&A 351, 827 (Paper II)
%
\bibitem[1999b]{Ekholm99b}
Ekholm,~T., Teerikorpi,~P., Theureau,~G. et al., 1999b,
A\&A 347, 99
%
\bibitem[1998]{Federspiel98}
Federspiel,~M., Tammann,~G.~A., Sandage,~A., 1998, ApJ 495, 115
%
\bibitem[1946]{Gamow46}
Gamow,~G., 1946, Nature 379, 549
%
\bibitem[1999]{Gibson99}
Gibson,~B.~K., Stetson,~P.~B., Freedman,~W.~L. et al., 1999,
ApJ, in press (astro-ph/9908149)
%
\bibitem[1969]{Gouguenheim69}
Gouguenheim,~L., 1969, A\&A 3, 281
%
\bibitem[1988]{Guhathakurta88}
Guhathakurta,~P., van Gorkum,~J.~H., Kotanyi,~C.~G.,
Balkowski,~C., 1988, AJ 96, 851
%
%
\bibitem[1980]{Hoffman80}
Hoffman,~G.~L., Olson,~D.~W., Salpeter,~E.~E., 1980, ApJ 242, 861
%
\bibitem[1956]{Humason56}
Humason,~M.~L., Mayall,~N.~U., Sandage,~A., 1956, AJ 61, 97
%
\bibitem[1999]{Lanoix99}
Lanoix,~P., 1999, PhD Thesis, University of Lyon 1
%
\bibitem[1999a]{Lanoix99a}
Lanoix,~P., Paturel,~G., Garnier,~R., 1999a, 
MNRAS 308, 969
%
\bibitem[1999b]{Lanoix99b}
Lanoix,~P., Garnier,~R., Paturel,~G. et al., 1999b,
Astron. Nach., 320, 21
%
\bibitem[1999c]{Lanoix99c}
Lanoix,~P., Paturel,~G., Garnier,~R., 1999c, ApJ 516, 188
%
\bibitem[1952]{Ogorodnikov52}
Ogorodnikov,~K.~F., 1952, Problems of Cosmogony 1, 150
%
\bibitem[1979]{Olson79}
Olson,~D.~W., Silk,~J., 1979, ApJ 233, 395
%
\bibitem[1976]{Peebles76}
Peebles,~J., 1976, ApJ 205, 318
%
%
\bibitem[1951]{Rubin51}
Rubin,~V.~C., 1951, AJ 56, 47
%
\bibitem[1988]{Rubin88}
Rubin,~V.~C., 1988 in {\sl World of Galaxies},
eds. H.~C.~Corwin and L.~Bottinelli,
New York: Springer, 431
%
\bibitem[1974]{Silk74}
Silk,~J., 1974, ApJ 193, 525
%
%
\bibitem[1997]{Teerikorpi97}
Teerikorpi,~P., 1997, ARA\&A 35, 101
%
\bibitem[1992]{Teerikorpi92}
Teerikorpi,~P., Bottinelli,~L., Gouguenheim,~L., Paturel,~G.,
1992, A\&A 260, 17 (Paper I)
%
\bibitem[1997]{Theureau97}
Theureau,~G., Hanski,~M., Ekholm,~T. et al., 1997,
A\&A 322, 730
%
\bibitem[1999]{Tittley99}
Tittley,~E.~R., Couchman,~H.~M.~P., 1999, astro-ph/9911365
\bibitem[1934]{Tolman34}
Tolman,~R.~C., 1934, Proc. Nat. Acad. Sci (Wash), 20, 169
%
\bibitem[1977]{Tully77}
Tully,~R.~B., Fisher,~J.~R., 1977, A\&A 54, 661
%
\bibitem[1982]{Tully82}
Tully,~R.~B., 1982, ApJ 257, 389
%
\bibitem[1984]{Tully84}
Tully,~R.~B., Shaya,~E.~J., 1984, ApJ 281, 31
%
\bibitem[1953]{deVaucouleurs53}
Vaucouleurs,~G.~de, 1953, AJ 58, 30
%
\bibitem[1958]{deVaucouleurs58}
Vaucouleurs,~ G.~de, 1958, AJ 63, 253
%
\bibitem[1976]{deVaucouleurs76}
Vaucouleurs,~G.~de, Vaucouleurs,~A.~de, Corwin,~H.~G.,
1976, Second Reference Catalogue of Bright Galaxies,
University of Texas Press, Austin (RC2)
%
\bibitem[1991]{deVaucouleurs91}
Vaucouleurs,~G.~de, Vaucouleurs,~A.~de, Corwin,~H.~G. et al,
1991, Third Reference Catalogue of Bright Galaxies,
Springer-Verlag (RC3)
\bibitem[1977]{Yahil77}
Yahil,~A., Tammann,~G.~A., Sandage,~A., 1977,
ApJ 217, 903
%
\end{thebibliography}
\end{document}